\documentclass[journal]{IEEEtran}
\usepackage{amsbsy,amsmath,amssymb,epsfig,bbm,mathrsfs,fancyhdr,fancyvrb,subfigure,url,cite,multirow,array}
\usepackage{graphicx,bm,makecell,utfsym}
\usepackage{epstopdf}
\usepackage{setspace}
\usepackage[ruled,linesnumbered]{algorithm2e}
\allowdisplaybreaks

\usepackage{textcomp,booktabs}
\usepackage[table,xcdraw]{xcolor}

\newcommand{\beq}{\begin{equation}}
\newcommand{\eeq}{\end{equation}}
\newcommand{\beqn}{\begin{eqnarray}}
\newcommand{\eeqn}{\end{eqnarray}}

\newtheorem{thm}{\protect\theoremname}

\newtheorem{rem}{\protect\remarkname}

\newtheorem{myDef}{\protect\Definition}
\providecommand{\theoremname}{\textbf{Theorem}}
\providecommand{\propositionname}{\textbf{Proposition}}
\providecommand{\remarkname}{\textbf{Remark}}
\providecommand{\lemmaname}{\textbf{Lemma}}
\providecommand{\corollaryname}{\textbf{Corollary}}
\providecommand{\Definition}{\textbf{Definition}}

\begin{document}

\title{Barycentric Coded Distributed Computing with Flexible Recovery Threshold for Collaborative Mobile Edge Computing}
\author{Houming Qiu, Kun Zhu,~\IEEEmembership{Member,~IEEE}, Dusit Niyato,~\IEEEmembership{Fellow,~IEEE}, Nguyen Cong Luong, \\Changyan Yi,~\IEEEmembership{Senior~Member,~IEEE} and Chen Dai
\thanks{H. Qiu is with the School of Software and IoT Engineering, Jiangxi University of Finance and Economics, Jiangxi 330032, China (email: qiuhouming@jxufe.edu.cn).}
\thanks{K. Zhu and C. Yi are with the College of Computer Science and Technology, Nanjing University of Aeronautics and Astronautics, Nanjing 210016, China (email: zhukun@nuaa.edu.cn; changyan.yi@nuaa.edu.cn).}
\thanks{D. Niyato is with the College of Computing and Data Science, Nanyang Technological University, Singapore 639798 (email: dniyato@ntu.edu.sg).}
\thanks{N. C. Luong is with the Faculty of Computer Science, Phenikaa University, Hanoi 12116, Vietnam (email: luong.nguyencong@phenikaa-uni.edu.vn).}
\thanks{C. Dai is School of Computer Science, Nanjing University of Posts and Telecommunications, Nanjing 210016, China (email: daichen@njupt.edu.cn).}
} \maketitle

\begin{abstract}
Collaborative mobile edge computing (MEC) has emerged as a promising paradigm to enable low-capability edge nodes to cooperatively execute computation-intensive tasks. However, straggling edge nodes (stragglers) significantly degrade the performance of MEC systems by prolonging computation latency. While coded distributed computing (CDC) as an effective technique is widely adopted to mitigate straggler effects, existing CDC schemes exhibit two critical limitations: (i) They cannot successfully decode the final result unless the number of received results reaches a fixed recovery threshold, which seriously restricts their flexibility; (ii) They suffer from inherent poles in their encoding/decoding functions, leading to decoding inaccuracies and numerical instability in the computational results. To address these limitations, this paper proposes an approximated CDC scheme based on barycentric rational interpolation. The proposed CDC scheme offers several outstanding advantages. Firstly, it can decode the final result leveraging any returned results from workers. Secondly, it supports computations over both finite and real fields while ensuring numerical stability. Thirdly, its encoding/decoding functions are free of poles, which not only enhances approximation accuracy but also achieves flexible accuracy tuning. Fourthly, it integrates a novel BRI-based gradient coding algorithm accelerating the training process while providing robustness against stragglers. Finally, experimental results reveal that the proposed scheme is superior to existing CDC schemes in both waiting time and approximate accuracy.
\end{abstract}


\section{Introduction}
\IEEEPARstart{W}{ith} the rapid advancements in mobile Internet, artificial intelligence (AI) and 5G beyond communication technologies, ubiquitous mobile devices such as smartphones, wearable devices, and smart home devices deliver various services to enhance convenience and quality of daily life. These services, including facial recognition~\cite{he2024enhancing}, augmented/virtual reality~\cite{chi2025h,zhu2025fpselector}, and AI-generated content~\cite{zhong2025generative}, generate a substantial number of delay-sensitive and computation-intensive tasks. Collaborative mobile edge computing (MEC) has gained widespread adoption since it offers computing, storage and network services for mobile devices at the network edge to meet ultra-low latency and high-reliability requirements~\cite{cai2024prioritized,li2025joint,li2019online,yi2019multi,cao2017distributed}. However, straggling edge nodes (stragglers) are inevitable in the collaborative MEC framework due to limited computing resources, network congestion, and constrained bandwidth. Typically, stragglers execute subtasks significantly more slowly than normal edge nodes or fail to complete them entirely. This unpredictably increases the overall task latency, thus becoming a critical performance bottleneck in collaborative MEC systems~\cite{lin2024coded,qiu2024resilient,ng2021comprehensive}.

To overcome the performance bottleneck caused by stragglers, coded distributed computing (CDC) is proposed to strengthen resilience against straggler effects for collaborative MEC systems. The core idea of CDC technology is to inject computational redundancy into each subtask, thereby enabling the mobile devices to decode the final desired computational result from partial sub-results returned by a subset of edge nodes~\cite{lu2025quantum}. In~\cite{lee2018speeding}, the authors were the first to address the issue of stragglers by proposing a CDC scheme for distributed matrix multiplication (DMM) that exploits maximum distance separable (MDS) codes. Considering the same issue, the authors in~\cite{yu2017polynomial} proposed the polynomial codes, which achieve lower communication costs. Correspondingly, the authors in~\cite{dutta2020optimal} proposed the MatDot codes, which have a lower recovery threshold compared to the polynomial codes. However, the MatDot codes incur higher communication and computation loads than the polynomial codes. In~\cite{yu2020straggler}, the authors proposed the entangled polynomial (EP) code for DMM tasks in distributed systems with stragglers. The proposed CDC schemes of~\cite{lee2018speeding,yu2017polynomial,dutta2020optimal,yu2020straggler} partition matrices in various ways and then design encoding functions to achieve the tolerance of stragglers. The related works are also found in~\cite{hasirciouglu2021bivariate,li2022private,song2023optimal,qiu2024secure,fahim2021numerically}.

Although these CDC schemes can mitigate the impact of stragglers, their recovery thresholds remain relatively high. To further reduce the recovery threshold, the authors in~\cite{jeong2021e} considered the issue of DMM from an approximate computational perspective and proposed the $\epsilon$-approximate MatDot code. The $\epsilon$-approximate MatDot code reduces the recovery threshold of the EP code~\cite{yu2020straggler} from $p^2q+q-1$ to $p^2q$ when the multiplication of matrices $\mathbf{A}$ and $\mathbf{B}$ are evenly partitioned into $p\times q$ and $q\times p$ submatrices. Inspired by~\cite{kiani2022successive}, the authors proposed a successive approximation coding (SAC) scheme, which achieves a tradeoff between execution time and accuracy. In addition, the authors in~\cite{jahani2021codedsketch} proposed a new encoding scheme, namely CodedSketch, which integrates count-sketch into the EP code~\cite{yu2020straggler} to reduce the recovery threshold. In~\cite{charalambides2021approximate}, the authors proposed an approximate weighted CR scheme for DMM tasks, which integrates the low-rank approximation of matrix products with distributed computing systems. However, these CDC schemes still require collecting an adequate number of returned results to decode the final result.

As aforementioned, several critical issues still limit the computation performance of CDC systems. Firstly, existing CDC schemes require a fixed number of computational results to recover the final result. This potentially prolongs the overall result waiting time due to the fixed recovery threshold. Secondly, the encoding/decoding functions in these CDC schemes contain intrinsic polarities, which lead to decoding errors and numerical instability in the computed results. Thirdly, existing works mainly focus on designing CDC scheme for a specific class of computational tasks, such as a DMM task. It is crucial to extend the consideration to a broader class of computational tasks, including arbitrary polynomial functions. In response to these issues, we design a novel approximated coding technique, referred to as the barycentric rational interpolation (BRI) code.
Our contributions in this paper are as follows:
\begin{itemize}
\item [$\bullet$]
Based on the collaborative MEC system, we propose a flexible BRI code using barycentric rational interpolation~\cite{floater2007barycentric}, which can decode the final result leveraging any returned results from workers. The final result becomes more accurate as more returned results are collected. Evidently, the BRI code guarantees a flexible recovery threshold. In particular, the most recent coding scheme, BACC~\cite{jahani2023berrut}, is a special instance of our proposed BRI code, with a detailed explanation provided in \textbf{\textit{Remark}}~\ref{rem3}.
\item [$\bullet$]
The BRI code achieves computations over both finite and real fields while maintaining numerical stability. We design the encoding and decoding functions based on barycentric rational interpolation~\cite{floater2007barycentric}, which is free of poles and supports arbitrarily high approximation orders. This property ensures that the decoding function remains free of singularities over both finite and real fields, while supporting arbitrarily distributed interpolation points without sacrificing approximation accuracy.
\item [$\bullet$]
We design the BRI code for a broad range of computational tasks, such as arbitrary polynomial functions. We propose a gradient coding algorithm based on the BRI code to overcome the straggler effects in distributed linear regression (LR). The experimental results demonstrate that the proposed algorithm is effective in tolerating stragglers and reducing training time.
\item [$\bullet$]
Finally, we conduct extensive experiments to validate the outstanding performance of our BRI code. The BRI code exhibits significantly shorter waiting times compared to baseline CDC schemes and achieves higher approximation orders.
\end{itemize}

\begin{table}
\center
\small
\renewcommand{\arraystretch}{1.2}
\caption{Major notations}
\label{table1}
\begin{tabular}{p{55pt}p{163pt}}
\toprule
Symbol   & Description \\
\midrule
$f$ & Arbitrary polynomial function\\
$k$ & Number of returned results \\
$\mathcal{K}$ & Set of indices of returned workers \\
$m$ & Number of submatrices\\
$\mathcal{M}$ & $\mathcal{M}\triangleq\{0,1,\ldots,m\}$\\
$N$ & Number of workers\\
$\mathcal{N}$ & Set of indices of $N$ workers \\
$n$ & Number of interpolation points\\
$s$ & Number of rows in matrix $\mathbf{X}$\\
$S$ & Number of straggers\\
$\mathcal{S}$ & Set of indexes of straggers\\
$t$ & Number of columns in matrix $\mathbf{X}$\\
$W_i$  & Worker node $i$\\
$\mathbf{X}$  & High-dimensional input matrix\\
$\tilde{\mathbf{X}}_i$  & Encoded matrix $i$\\
$\mathbf{X}_i$  & Submatrix $i$\\
$\mathbf{Y}$  & Task result\\
$\tilde{\mathbf{Y}}_i$  & Sub-result from worker $i$\\
\bottomrule
\end{tabular}
\end{table}

The rest of the paper is organized as follows. In Section~\uppercase\expandafter{\romannumeral2}, we introduce the related works. Then, the system model and the current critical problems are presented in Section~\uppercase\expandafter{\romannumeral3}. In Section~\uppercase\expandafter{\romannumeral4}, we introduce our proposed scheme and provide three illustrating examples. Then, we propose a BRI-based gradient coding algorithm in Section~\uppercase\expandafter{\romannumeral5}. In Section~\uppercase\expandafter{\romannumeral6}, we provide a detailed theoretical analysis. After that, we provide extensive experiments for the proposed scheme and baseline schemes in Section~\uppercase\expandafter{\romannumeral7}. Finally, in Section \uppercase\expandafter{\romannumeral8}, we conclude this work.


For the convenience, the major notations of this paper is detailed in Table~\ref{table1}.

\begin{table*}
\centering
\footnotesize
\renewcommand{\arraystretch}{1.5}
\caption{Comparison of existing CDC schemes}
\label{tbCompCDC}
\begin{tabular}{|m{3.8cm}<{\centering}|m{1.5cm}<{\centering}|m{1.4cm}<{\centering}|m{2cm}<{\centering}|m{2cm}<{\centering}|m{2.1cm}<{\centering}|m{2cm}<{\centering}|}
\hline
Refs. & Support finite fields& Support real fields & Support matrix-vector multiplication task & Support matrix-matrix multiplication task & Support arbitrary polynomial functions&Support variable recovery thresholds\\
\hline
\cite{qiu2024coded,dutta2019short,wang2019computation,wang2021batch}& $\usym{1F5F8}$ & $\usym{2613}$ & $\usym{1F5F8}$ & $\usym{2613}$ & $\usym{2613}$ & $\usym{2613}$\\
\hline
\cite{lee2018speeding} & $\usym{1F5F8}$ & $\usym{2613}$ & $\usym{1F5F8}$ & $\usym{1F5F8}$ & $\usym{2613}$ & $\usym{2613}$\\
\hline
\cite{yu2019lagrange} & $\usym{1F5F8}$ & $\usym{2613}$ & $\usym{1F5F8}$ & $\usym{1F5F8}$ & $\usym{1F5F8}$ & $\usym{2613}$\\
\hline
\cite{asheralieva2021fast,soleymani2021analog} & $\usym{1F5F8}$ & $\usym{1F5F8}$ & $\usym{1F5F8}$ & $\usym{1F5F8}$ & $\usym{1F5F8}$ & $\usym{2613}$\\
\hline
\cite{makkonen2024general,khalesi2023multi,yu2017polynomial,dutta2020optimal,li2022private,song2023optimal,qiu2024secure,soto2022rook,tauz2022variable,yang2019secure,bitar2022adaptive}&$\usym{1F5F8}$ & $\usym{2613}$ & $\usym{2613}$ & $\usym{1F5F8}$ & $\usym{2613}$ & $\usym{2613}$ \\
\hline
\cite{lin2024coded,yu2020straggler,hasirciouglu2021bivariate,fahim2021numerically,jeong2021e,kiani2022successive,jahani2021codedsketch,charalambides2021approximate,jia2021cross}&$\usym{1F5F8}$ & $\usym{1F5F8}$ & $\usym{2613}$ & $\usym{1F5F8}$ & $\usym{2613}$ & $\usym{2613}$ \\
\hline
\cite{Das2022Coded,Das2023Distributed,Das2023Distributed2,das2022unified}&  $\usym{1F5F8}$ &$\usym{1F5F8}$ & $\usym{1F5F8}$ & $\usym{1F5F8}$ & $\usym{2613}$ & $\usym{2613}$\\
\hline
BRI code (our scheme) &  $\usym{1F5F8}$ &$\usym{1F5F8}$ & $\usym{1F5F8}$ & $\usym{1F5F8}$ & $\usym{1F5F8}$ & $\usym{1F5F8}$\\
\hline
\end{tabular}
\end{table*}

\section{Related Work}
\subsection{Coded Distributed Computing}
In recently year, CDC has gained widespread attention for enabling distributed computing systems to tolerate the presence of stragglers~\cite{qiu2025approximated,cohen2021stream,kakar2021codes}. Considerable effort is being devoted to designing CDC schemes for efficient distributed computing tasks, such as matrix-matrix multiplication~\cite{lee2018speeding,yu2017polynomial,dutta2020optimal,yu2020straggler,hasirciouglu2021bivariate,li2022private,song2023optimal,jeong2021e,kiani2022successive,jahani2021codedsketch,charalambides2021approximate,qiu2024secure}.
On the other hand, numerous CDC schemes~\cite{qiu2024coded,dutta2019short,wang2019computation,wang2021batch,Das2022Coded,Das2023Distributed,Das2023Distributed2} are designed for high-dimensional matrix-vector multiplications.

In~\cite{dutta2019short}, the authors proposed a Short-Dot scheme to alleviate the effects of straggling workers. This scheme achieves shorter dot
product computations by enforcing sparsity on encoded submatrices. Nonetheless, the recovery threshold is larger when the dot product length is
shorter. To overcome this drawback, the authors in~\cite{wang2019computation} proposed an $s$-diagonal code, which lowers both the recovery threshold
and computational cost compared to the Short-Dot~\cite{dutta2019short}. The $s$-diagonal code leverages the diagonal structure to construct a
decoding method. Despite these advantages, the task waiting time of the $s$-diagonal code remains unsatisfactory.

To reduce the task waiting time, the authors in \cite{wang2021batch} proposed the BPCC scheme, which employs a batch task allocation approach to
fully utilize the computational capacity of each workers, thereby accelerating the decoding process. However, the encoding matrix required by the
BPCC scheme entails high computational costs. In~\cite{Das2022Coded}, the authors proposed a $\beta$-level coding scheme based on combinatorial
designs, which fully utilizes the work already done by each worker while considering sparse matrices. The $\beta$-level coding scheme improves the
recovery threshold and reduces the waiting time. However, the $\beta$-level encoding scheme cannot fully meet the optimal recovery threshold. To
improve the $\beta$-level encoding scheme, the authors in~\cite{Das2023Distributed,Das2023Distributed2} proposed an innovative CDC scheme in which
the encoding matrices allocated to each worker are generated by randomly combining a small subset of submatrices. In~\cite{das2022unified}, the authors proposed an efficient CDC scheme for distributed sparse matrix multiplication that achieves the optimal recovery threshold while substantially accelerating sparse matrix computation. It can not only leverage partial results from straggling nodes but also exploit the weight of the coding scheme to trade off the recovery threshold.

Since these works only consider a specific computational task, such as matrix multiplication, the application scope of existing CDC schemes is limited. To overcome this limitation, the authors in~\cite{yu2019lagrange} proposed the LCC scheme, which is developed for computing arbitrary polynomial functions over distributed systems. However, the LCC scheme  is only applicable to finite fields. In~\cite{soleymani2021analog}, the authors extended the LCC scheme~\cite{yu2019lagrange} from the finite fields to the analog domain, proposing an analog LCC (ALCC) scheme. In addition, the authors in~\cite{soto2022rook,jia2021cross,tauz2022variable} also considered the same issue. However, the computational costs of these CDC schemes are prohibitively high. For instance, the encoding and decoding functions of the LCC and ALCC schemes involve numerous multiplication operations. While matrix blocks increase linearly, the associated multiplication operations rise exponentially. As compared to CDC schemes, we aim to design an encoding scheme that has low computational cost and is applicable to arbitrary polynomial functions.

The aforementioned CDC schemes focus on reducing the recovery threshold to enhance resilience against stragglers. However, these encoding schemes can only decode the final result when the number of received results reaches a fixed value, known as the recovery threshold. This potentially prolongs the task waiting times. As compared to these CDC schemes, we aim to design an innovative CDC scheme, which provides a flexible decoding strategy. The final result can be reconstructed utilizing any of the returned computational results. In~\cite{jahani2023berrut} and \cite{qiu2023secure}, the authors addressed this issue by proposing the BACC scheme and the SPACDC scheme, respectively. However, both the BACC and SPACDC schemes exhibit relatively low approximation rates. This motivates us to consider enhancing the approximation rates for the proposed coding scheme.

Furthermore, we compare the BRI code with the existing CDC schemes in Table~\ref{tbCompCDC}. As shown in Table~\ref{tbCompCDC}, our BRI code is superior to the existing CDC schemes in various respects. For instance, the BRI code is applicable not only to matrix multiplication tasks, but also to arbitrary polynomial functions where the variable is a matrix~\cite{qiu2025approximated,jahani2023berrut}. In particular, the BRI code achieves a flexible recovery threshold.

\subsection{Coded Distributed Computing in MEC}
In unreliable MEC systems, coded distributed computing has attracted considerable attention due to its ability to provide robustness against stragglers~\cite{asheralieva2021fast,qiu2024coded,schlegel2022privacy}. In~\cite{asheralieva2021fast}, the authors proposed a Lagrange coded computing (LCC)-based framework for secure and low-latency task offloading in MEC systems. However, the encoding and decoding latency of the LCC scheme is a major performance bottleneck of the framework. To further reduce the overall latency, the authors in~\cite{qiu2024coded} proposed a novel CDC scheme based on a distributed edge computing system that exploits the work done by stragglers. For the same problem, the authors in~\cite{schlegel2022privacy} proposed a fast and secure CDC scheme that aims to minimize the overall latency for executing matrix multiplications in a collaborative MEC scenario. Although the aforementioned CDC schemes effectively provide robustness against stragglers, their performance remains constrained by a fixed recovery threshold.

\section{System Model and problem formulation}
In this section, we introduce the system model and some related definitions. Then, we present the current critical issues for approximated code computing in a collaborative MEC framework with stragglers.
\subsection{Related Definitions}

\begin{myDef}
(\textbf{Stragglers}~\cite{qiu2024secure})~Stragglers refer to some workers that return results to the master much more slowly than other workers or fail to complete their computing tasks. The primary reason stems from the heterogeneity of workers in terms of computational power, storage resources, and network bandwidth. In addition, the occurrence of straggling nodes may result from unexpected factors, such as device malfunctions.
\end{myDef}

\begin{myDef}
(\textbf{Recovery Threshold}~\cite{qiu2024resilient})~The recovery threshold is defined as the minimum number of computed results from workers that can correctly decode to obtain the final result.
\end{myDef}

\begin{myDef}
(\textbf{Barycentric Rational Interpolation~\cite{berrut2004barycentric,jahani2023berrut,floater2007barycentric}})~Given a set of $n+1$ distinct points $a\leq x_0<x_1<\cdots<x_{n}\leq b$, and the corresponding function values $y_i=f(x_i),i\in\{0,1,\ldots,n\}$, where $a$ and $b$ are two real numbers. Barycentric rational interpolation finds a rational function $v(x)$ to approximate the given data points. The rational function $v(x)$ is given by
\begin{equation}
\begin{split}
\label{eqvx}
v(x)=\frac{\sum_{j=0}^{n}\frac{w_jy_j}{x-x_j}}{\sum_{i=0}^{n}\frac{w_i}{x-x_i}},
\end{split}
\end{equation}
where $w_j, j \in \{0,1,\ldots,n \}$ are the weights defined as
\begin{equation}
\begin{split}
w_j=\frac{1}{\prod_{0<k<n,k\neq j}(x_j-x_k)}.
\end{split}
\end{equation}
\end{myDef}

\begin{figure}[!t]
\centering
\includegraphics[width=3.3in]{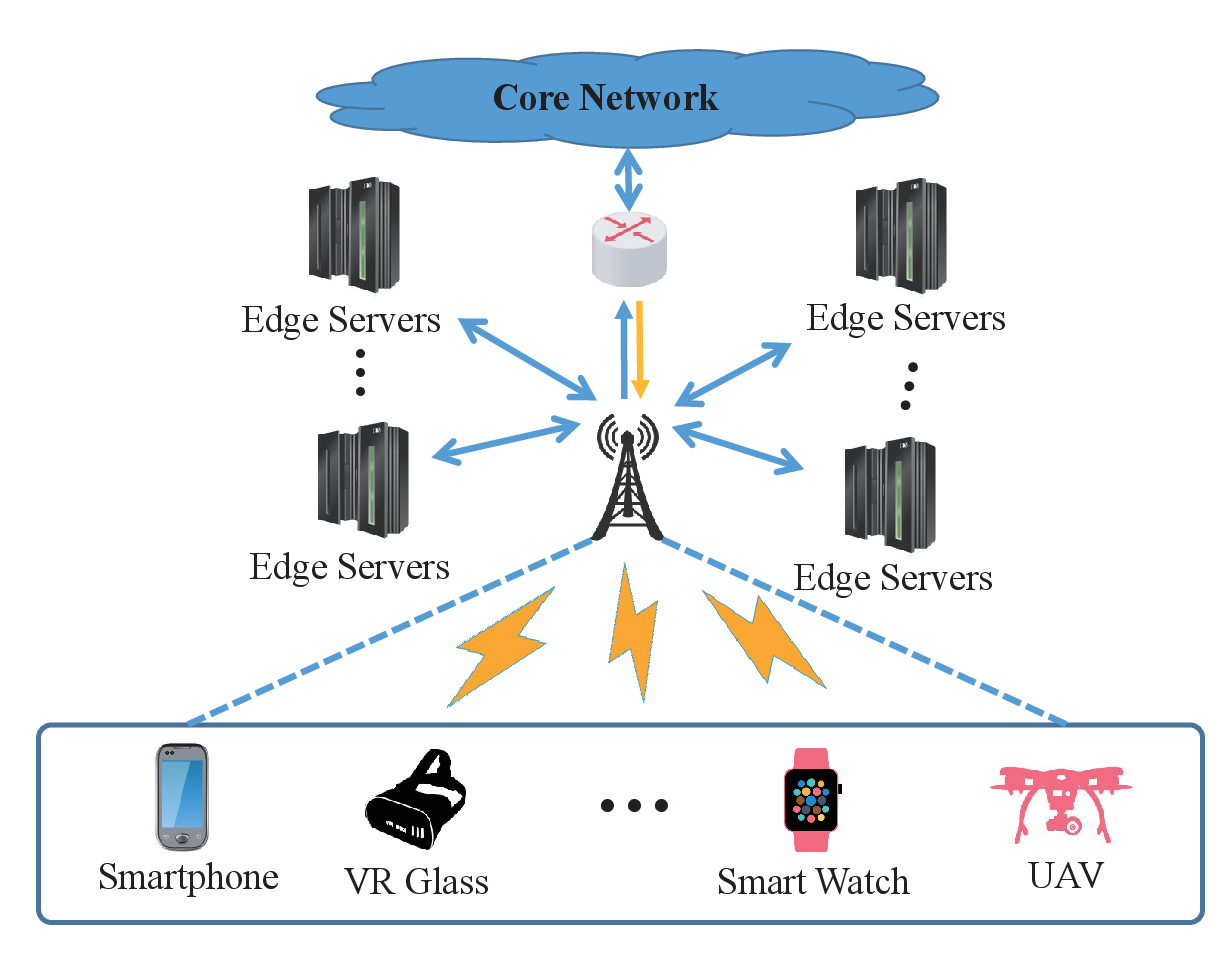}
\caption{Overview of the system framework for the collaborative mobile edge computing.}
\label{fig:systemFramework}
\end{figure}

\begin{figure}[!t]
\centering
\includegraphics[width=3.3in]{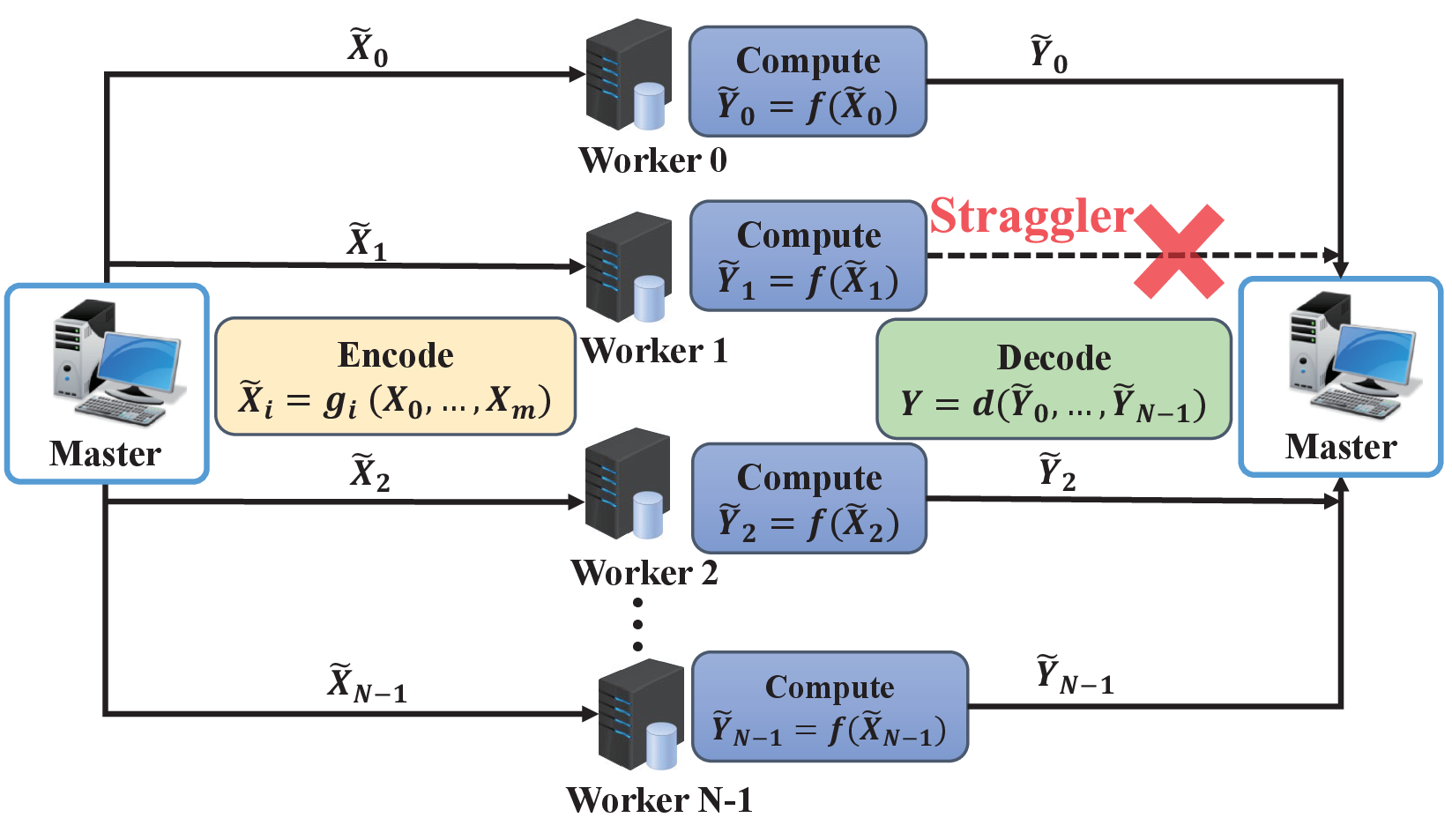}
\caption{An illustration of using a flexible barycentric approximated CDC (BRI) scheme for calculating a polynomial function $f$ over a high-dimensional matrix $\mathbf{X}=(\mathbf{X}_0, \ldots, \mathbf{X}_m)$ within a collaborative MEC system. In the system, there are $S$ stragglers among $N$ workers. Each worker executes its computing task and returns the computed result to the master. Finally, the master can obtain an approximation of the final result from any returned computational results.}
\label{fig:systemModel}
\end{figure}

\subsection{System Model}
We consider a collaborative MEC system is composed of (i) local mobile devices (e.g., smartphones, smart watches, VR glasses, and UAVs) with constrained computing and storage resources, and (ii) multiple edge servers that provide computational services to nearby mobile devices. As shown in Fig.~\ref{fig:systemFramework}, each mobile device acts as a master distributing numerous delay-sensitive and computation-intensive tasks to nearby edge servers. The edge servers play the roles of workers collaborating with each other to complete tasks assigned by the master. We consider that all tasks are computation-intensive and edge servers possess the computational capabilities required to execute the tasks assigned to them by mobile devices. In the system, each mobile device may become unavailable during task execution due to factors such as mobility and battery capacity. For brevity, we designate each mobile device as a master and edge devices as workers. Each worker are heterogeneous in their capabilities. Moreover, worker $i\in\mathcal{N}$ is denoted by $W_i$, where $\mathcal{N}\triangleq\{0,1,\ldots, N-1\}$. We assume that there are $S$ stragglers among $N$ workers in the system. The master aims to obtain an approximation of evaluating an arbitrary polynomial function $f:\mathbb{V}\rightarrow\mathbb{U}$ over an input dataset $\mathbf{X}=(\mathbf{X}_0, \mathbf{X}_1, \ldots, \mathbf{X}_m)$, i.e., $\mathbf{Y}\approx f(\mathbf{X}_i)$ for $i\in\{0,1,\ldots, m\}$. $\mathbb{V}$ and $\mathbb{U}$ denote two real matrix spaces. Each element $\mathbf{X}_i$ of the dataset $\mathbf{X}$ has a dimension of $s\times t$, where $s$ and $t$ are two positive integers. $\mathbb{F}$ denotes a sufficiently large field.

Figure~\ref{fig:systemModel} illustrates the CDC process in the collaborative MEC system, which operates as follows:

\begin{enumerate}
\item
\emph{\textbf{Data Encoding:}}
The master firstly encodes the input matrix by encoding function as follows:
\begin{equation*}
\begin{split}
\tilde{\mathbf{X}}_i=\tilde{\mathrm{g}}_i(\mathbf{X})~\text{for}~i\in\mathcal{N},
\end{split}
\end{equation*}
where $\mathcal{N}\triangleq\{0,1,\ldots,N-1\}$. $\tilde{\mathbf{X}}_i~(i\in\mathcal{N})$ is the encoded data. The master sends $\tilde{\mathbf{X}}_i$ to worker $W_i$ for $i\in\mathcal{N}$.
\item
\emph{\textbf{Task Computing:}}
Worker $W_i$ receives encoded data $\tilde{\mathbf{X}}_i$ from the master. Then, worker $W_i$ executes an assigned computational task $\tilde{\mathbf{Y}}_i=f(\tilde{\mathbf{X}}_i)$. The worker returns the computational result to the master immediately after completing the task. It is noteworthy that the stragglers may fail to execute their tasks or return the computational results much more slowly than other workers.

\item
\emph{\textbf{Result Decoding:}}
The master waits and receives the computational results returned by the workers. Then, the master decodes the final result $\mathbf{Y}$ via the following decoding function:
\begin{equation}
\begin{split}
\mathbf{Y}=\hslash_{\mathcal{K}}\bigg(\{\tilde{\mathbf{Y}}_i\}_{i\in\mathcal{K}}\bigg),
\label{eqDefunc}
\end{split}
\end{equation}
where $\mathcal{K}$ is the set of the indices of the fastest $k=|\mathcal{K}|$ workers. $\hslash_{\mathcal{K}}$ is a decoding function.
\end{enumerate}

\subsection{Problem Description}
This paper studies the approximated computing for a polynomial function $f$ over a high-dimensional dataset $\mathbf{X}$ within a collaborative MEC system with a master and $N$ workers. We aim to design an approximated coded distributed computing scheme over real fields, which can decode the final result by utilizing any returned computational results, i.e., Eq.~\eqref{eqDefunc}. In other words, the proposed CDC scheme relaxes the constraints on the recovery threshold. Hence, the waiting time for an assigned task is greatly reduced.
Moreover, we aim to address the issues of approximation error and numerical instability by designing the encoding and decoding functions based on barycentric rational interpolation, which is free of poles and supports arbitrarily high approximation orders. This feature guarantees that the denominator of the decoding function remains non-zero over both finite and real fields, thereby preventing the occurrence of poles and the associated degradation in decoding accuracy.

\section{The Proposed Coding Technique}
In this section, we introduce the proposed CDC technique, termed the BRI code. In particular, the BRI code achieves a flexible recovery threshold.

The master aims to approximately evaluate arbitrary polynomial function $f$ over an input high-dimensional matrix $\mathbf{X}=(\mathbf{X}_0, \mathbf{X}_1, \ldots, \mathbf{X}_m)$, where each element $\mathbf{X}_i$ for $i\in\{0,1,\ldots, m\}$ has a dimension of $s\times t$. The distributed computing system is composed of a master and $N$ workers, where the number of stragglers is $S$.
The BRI code consists of three phases: data encoding, task computing, and result decoding. Next, we introduce the BRI code in accordance with these three phases.
\subsection{\textbf{Data Encoding}}
Firstly, the master generates $N$ encoded matrices by encoding the input matrices $\{\mathbf{X}_{0},\mathbf{X}_{1},\ldots,\mathbf{X}_{m}\}$ using the following encoding function based on barycentric rational interpolation~\cite{floater2007barycentric}:
\begin{equation}
\begin{split}
r(x)=\frac{\sum_{i=0}^{m-d}\phi_i(x)l_i(x)}{\sum_{j=0}^{m-d}\phi_j(x)},
\label{eqEdfunc}
\end{split}
\end{equation}
where $d$ is an any integer satisfying $0\leq d\leq m$. $\phi_i(x)$ and $l_i(x)$ are as follows:
\begin{equation}
\begin{split}
\phi_i(x)=\prod_{j=0}^{i-1}(x-\alpha_j)\prod_{k=i+d+1}^{m}(\alpha_k-x),
\label{eqEdfun1}
\end{split}
\end{equation}
and
\begin{equation}
\begin{split}
l_i(x)=\sum_{k=i}^{i+d}\prod_{j=i,j\neq k}^{i+d}\frac{x-\alpha_j}{\alpha_k-\alpha_j}\mathbf{X}_k,
\label{eqEdfunc2}
\end{split}
\end{equation}
where $\alpha_0,\alpha_1,\ldots,\alpha_m$ are any $m+1$ distinct values from field $\mathbb{F}$. 
It is worth noting that $r(\alpha_i)=\mathbf{X}_i$ for $i\in\mathcal{M}\triangleq\{0,1,\ldots,m\}$.

Then, the master sends encoded matrices $\tilde{\mathbf{X}}_i=r(z_i)$ to worker $W_i$ for $i\in\mathcal{N}$. $z_0,z_1,\ldots,z_{N-1}$ are $N$ distinct values from field $\mathbb{F}$ while satisfying $\{\alpha_i\}^{m}_{i=0}\cup\{z_i\}_{i=0}^{N-1}=\varnothing$.

\subsection{\textbf{Task Computing}}
Each worker $W_i$ receives the encoded data $\tilde{\mathbf{X}}_i$ from the master, and then executes assigned computing task $\tilde{\mathbf{Y}}_i=f(\tilde{\mathbf{X}}_i)$. Worker $W_i$ for $i\in\mathcal{N}$ immediately returns the result $\tilde{\mathbf{Y}}_i$ to the master once the computational task is completed.

In the system, it is essential to highlight that not all workers return computational results to the master in a timely manner. Some workers, known as stragglers, may return computational results significantly more slowly than other workers or may even fail to return results altogether. Fortunately, the proposed BRI code effectively addresses this issue.

\subsection{\textbf{Result Decoding}}
The master waits and receives the returned computational result $\tilde{\mathbf{Y}}_i$ from worker $W_i$ for $i\in\mathcal{K}$. Then, the master designs a decoding function based on returned results. The decoding function is defined by
\begin{equation}
\begin{split}
h(x)=\frac{\sum_{i=0}^{k-d}\varphi_i(x)\zeta_i(x)}{\sum_{j=0}^{k-d}\varphi_j(x)},
\label{eqDedfunc}
\end{split}
\end{equation}
$\varphi_i(x)$ and $\zeta_i(x)$ are as follows:
\begin{equation}
\begin{split}
\varphi_i(x)=\prod_{j=0}^{i-1}(x-z_{k_j})\prod_{\tau=i+d+1}^{k}(z_{k_\tau}-x),
\end{split}
\end{equation}
and
\begin{equation}
\begin{split}
\zeta_i(x)=\sum_{\tau=i}^{i+d}\prod_{j=i,j\neq \tau}^{i+d}\frac{x-z_{k_j}}{z_{k_\tau}-z_{k_j}}\tilde{\mathbf{Y}}_{k_\tau},
\end{split}
\end{equation}
where $k=|\mathcal{K}|$ and $\mathcal{K}$ denotes the set of indices for the workers that return their results to the master. $k_i$ represents the $i$-th element of the $\mathcal{K}$. It is observed that Eq.~\eqref{eqDedfunc} is designed to approximately interpolate $f(r(x))$ using the interpolation points $(z_i,f(r(z_i))$ for $i\in\mathcal{K}$.

Subsequently, the master calculates the approximation results $\mathbf{Y}_i=f(\mathbf{X}_i)\approx h(\alpha_i)$ for $i\in\mathcal{M}$. The detailed procedure of the BRI code is outlined in Algorithm~\ref{algBRI}.

\begin{algorithm}[!t]
\label{algBRI}
\caption{BRI code}
\KwIn{$\mathbf{X}, N, S, m, s, t$}
\KwOut{$\{\mathbf{Y}_i\}_{i=0}^{m}$}
[~\uppercase\expandafter{\romannumeral1}~]~\textbf{Data Encoding:}~the master completes the encoding of $\mathbf{X}=(\mathbf{X}_0, \mathbf{X}_1, \ldots, \mathbf{X}_m)$\;
\For{$i=0:N-1$}
{
    The master obtain encoded matrices: $\tilde{\mathbf{X}}_i=r(z_i)$\;
    The master distributes $\tilde{\mathbf{X}}_i$ to worker $W_i$\;
}
[~\uppercase\expandafter{\romannumeral2}~]~\textbf{Task Computing:} $W_i$ executes task: $\tilde{\mathbf{Y}}_i=f(\tilde{\mathbf{X}}_i)$\;
\For{$i=0:N-1$}
{
    \If{$W_i$ \rm{receives} $\tilde{\mathbf{X}}_i$ \rm{from the master}}
    {
        $W_i$ executes $\tilde{\mathbf{Y}}_i=f(\tilde{\mathbf{X}}_i)$\;
        \If{$W_i$ \rm{successfully computes} $\tilde{\mathbf{Y}}_i$}
        {
            Worker $W_i$ returns $\tilde{\mathbf{Y}}_i$ to the master\;
        }
    }
}
[~\uppercase\expandafter{\romannumeral3}~]~\textbf{Result Decoding:} the master decodes $\mathbf{Y}$\;
The master adds the received index of the worker to the set $\bm{\mathcal{K}}$\;
The master constructs $(z_i,f(r(z_i))$ for $i\in\mathcal{K}$\;
The master designs the decoding function: $h(x)=\frac{\sum_{i=0}^{k-d}\varphi_i(x)\zeta_i(x)}{\sum_{j=0}^{k-d}\varphi_j(x)}$\;
The master executes $\mathbf{Y}_i=f(\mathbf{X}_i)\approx h(\alpha_i)$ for $i\in\mathcal{M}$\;
\textbf{Return:}{~$\mathbf{Y}_i$}
\end{algorithm}

\begin{rem}
Most existing coding schemes, such as~\cite{lee2018speeding,yu2017polynomial,dutta2020optimal,yu2020straggler,hasirciouglu2021bivariate,li2022private,song2023optimal,jeong2021e,kiani2022successive,jahani2021codedsketch,charalambides2021approximate,qiu2024secure,fahim2021numerically,qiu2024coded,dutta2019short,wang2019computation,wang2021batch,Das2022Coded,Das2023Distributed,Das2023Distributed2}, are unable to decode the final result unless the number of received results meets the fixed value, i.e., recovery threshold. Compared to these schemes, the BRI code provides a more flexible decoding scheme regarding the minimum number of required results. Particularly, the BRI code can utilize any returned computational results to obtain an approximation of the final result. Furthermore, increasing the number of returned results used for decoding improves the accuracy of the recovered final result. We validate this property via extensive experiments in Section~\uppercase\expandafter{\romannumeral5}. It is worth mentioning that the decoding results of the aforementioned CDC schemes are exact, whereas those obtained from our proposed coding scheme are approximate. Essentially, our scheme offers a trade-off between the recovery threshold and computational accuracy.
\end{rem}

\begin{rem}
By designing the encoding and decoding functions as shown in Eqs.~\eqref{eqEdfunc} and~\eqref{eqDedfunc}, it can be easily verified that the proposed BRI code is free of poles. Therefore, the BRI code guarantees numerical stability and offers more high approximations, as demonstrated in Theorem~\ref{thm0}. We experimentally demonstrate this characteristic in Section~\uppercase\expandafter{\romannumeral5} through comprehensive evaluations. Furthermore, the absence of poles in the denominator of the decoding function enhances the reliability of the BRI code in practical applications, making it suitable for a wider range of scenarios.
\end{rem}

\begin{rem}
\label{rem3}
The BRI code becomes the Berrut approximated coded computing (BACC) scheme~\cite{jahani2023berrut} when $d=0$. It indicates that the BACC scheme is a special instance of our proposed BRI code. The encoding function of the BACC scheme is given by
\begin{equation}
\begin{split}
h_{\rm{BACC}}(x)=\sum_{i=0}^{m}\frac{\frac{(-1)^i}{(x-\alpha_i)}}{\sum_{j=0}^{m}\frac{(-1)^j}{(x-\alpha_j)}}.
\label{eqDedfuncBACC}
\end{split}
\end{equation}
\end{rem}

\subsection{Illustrating Examples}
In this subsection, we provide three illustrative examples to elaborate on the key idea of our BRI code, the well-known LCC scheme, and EP code. Then, we compare the BRI code with the LCC scheme and EP code.

\subsubsection{\textbf{Example~\uppercase\expandafter{\romannumeral4}.1.}~(Our BRI code)}
In this example, we consider a specific large-scale computing task $f'(\mathbf{X})=\mathbf{X}^T\mathbf{X}$ in a CDC system, where $\mathbf{X}=(\mathbf{X}_0, \mathbf{X}_1, \mathbf{X}_2)$ and $\mathbf{X}_i\in\mathbb{F}^{s\times t}$ for $i=\{0,1,2\}$. Thus, the objective of the master is to calculate $f'(\mathbf{X})=\mathbf{X}_i^T\mathbf{X}_i$ for $i=\{0,1,2\}$. The system consists of a master and $N=9$ workers, where the number of stragglers is $S=1$. To simplify the problem, we set the system parameters as follows: $m=2$ and $d=0$.

Firstly, the encoding function is given by
\begin{equation}
\begin{split}
r'(x)&=\frac{1}{(x-\alpha_0')\kappa(x)}\mathbf{X}_0-\frac{1}{(x-\alpha_1')\kappa(x)}\mathbf{X}_1\\
&+\frac{1}{(x-\alpha_2')\kappa(x)}\mathbf{X}_2,
\label{eqEdfuncExam1}
\end{split}
\end{equation}
where $\kappa=1/(x-\alpha_0')-1/(x-\alpha_1')+1/(x-\alpha_2')$, $\alpha_0'$, $\alpha_1'$, and $\alpha_2'$ are any three distinct values from field $\mathbb{F}$.

Then, the master generates $N=9$ encoded matrices $\tilde{\mathbf{X}}_i'=r'(z_i')$, where $z_0',z_1',\ldots,z_8'$ are $9$ distinct values from field $\mathbb{F}$ while satisfying $\{\alpha_i'\}^{2}_{i=0}\cup\{z_i'\}_{i=0}^{8}=\varnothing$. In the next step, the master sends $\tilde{\mathbf{X}}_i'$ to worker $W_i$ for $i\in\{0,1,\ldots,8\}$.

After receiving the encoded matrix $\tilde{\mathbf{X}}_i'$ from the master, the worker executes its task $\tilde{\mathbf{Y}}_i'=f'(\tilde{\mathbf{X}}_i')$. The worker immediately returns the result $\tilde{\mathbf{Y}}_i'$ to the master after completing the computational task. On the other hand, the master waits and receives the returned computational result $\tilde{\mathbf{Y}}_i'$ for $i\in\mathcal{K}'$, where $\mathcal{K}'$ is the set of indices for the workers that return their results to the master. Then, the master constructs a decoding function based on the returned computational results as follows:
\begin{equation}
\begin{split}
h'(x)=\frac{\sum_{i=0}^{k'}\varphi_{k_i'}'(x)\zeta_{k_i'}'(x)}{\sum_{j=0}^{k'}\varphi_{k_j'}'(x)},
\label{eqDedfuncExam1}
\end{split}
\end{equation}
where $\varphi_i'(x)$ and $\zeta_i'(x)$ are as follows:
\begin{equation}
\begin{split}
\varphi_i'(x)=\prod_{j=0}^{i-1}(x-z_{k_j'}')\prod_{t=i'+1}^{k'}(z_{k_t'}'-x),
\end{split}
\end{equation}
and
\begin{equation}
\begin{split}
\zeta_i'(x)=\sum_{t=i}^{i}\prod_{j=i,j\neq t}^{i}\frac{x-z_{k_j'}'}{z_{k_t'}'-z_{k_j'}'}\tilde{\mathbf{Y}}_{k_t'}',
\end{split}
\end{equation}
where $k'=|\mathcal{K}'|$ and $k_i'$ represents the $i$-th element of the $\mathcal{K}'$. Consequently, the master obtains the approximation results $\mathbf{Y}_i'=f'(\mathbf{X}_i')\approx h'(\alpha_i')$ for $i\in\{0,1,2\}$.

\subsubsection{\textbf{Example~\uppercase\expandafter{\romannumeral4}.2.}~(LCC scheme~\cite{yu2019lagrange})}

In this example, the master performs the same computing task, i.e., $f'(\mathbf{X})=\mathbf{X}^T\mathbf{X}$ using the LCC scheme in a distributed system. For fair comparison, the system parameters are set as $N=9,S=1,A=0$, and $T=0$.

Similarly, the master firstly encodes the input dataset $\mathbf{X}_0,\mathbf{X}_1,\mathbf{X}_2$ using the following encoding function:
\begin{equation}
\begin{split}
r''(x)&=\frac{(x-\alpha_1'')(x-\alpha_2'')}{(\alpha_0''-\alpha_1'')(\alpha_0''-\alpha_2'')}\mathbf{X}_0+\frac{(x-\alpha_0'')(x-\alpha_2'')}{(\alpha_1''-\alpha_0'')(\alpha_1''-\alpha_2'')}\mathbf{X}_1\\
&+\frac{(x-\alpha_0'')(x-\alpha_1'')}{(\alpha_2''-\alpha_0'')(\alpha_2''-\alpha_1'')}\mathbf{X}_2,
\label{eqEdfuncExam2}
\end{split}
\end{equation}
where $\alpha_0''$, $\alpha_1''$, and $\alpha_2''$ are any three distinct values from the field $\mathbb{F}$.

Then, the master generates $N=9$ encoded matrices $\tilde{\mathbf{X}}_i''=r''(z_i'')$, where $z_0'',z_1'',\ldots,z_8''$ are $9$ distinct values from field $\mathbb{F}$ while satisfying $\{\alpha_i''\}^{2}_{i=0}\cup\{z_i''\}_{i=0}^{8}=\varnothing$. Following this, the master sends $\tilde{\mathbf{X}}_i''$ to worker $W_i$ for $i\in\{0,1,\ldots,8\}$. After receiving the encoded matrix $\tilde{\mathbf{X}}_i''$ from the master, worker $i$ performs the assigned task $\tilde{\mathbf{Y}}_i''=f'(\tilde{\mathbf{X}}_i'')$. The worker immediately returns the result $\tilde{\mathbf{Y}}_i''$ to the master after completing the computational task.

On the other hand, the master waits and receives the returned computational result $\tilde{\mathbf{Y}}_i''$ for $i\in\mathcal{K}''$, where $\mathcal{K}''$ is the set of indices for the workers that return their results to the master. Then, the master constructs the following decoding function based on the returned computational results:
\begin{equation}
\begin{split}
h''(x)=\sum_{j\in\mathcal{K}''}\tilde{\mathbf{Y}}_j''\prod_{i\in\mathcal{K}'',i\neq j}\frac{x-z_i''}{z_j''-z_i''},
\label{eqDedfuncExam2}
\end{split}
\end{equation}
where $k''=|\mathcal{K}''|$. Actually, Eq.~\eqref{eqDedfuncExam2} is designed to approximately interpolate $f'(r''(z))$ using the interpolation points $(z_i'',f(r(z_i''))$ for $i\in\mathcal{K}''$. The degree of the function $f'(r''(z))$ is $4$. Consequently, five interpolation points are required to successfully interpolate the function $f'(r''(z))$, i,e., $k''\geq 5$. Therefore, the recovery threshold of the LCC scheme is $5$ under this setting. For a more detailed description of the LCC scheme, please see Section~\uppercase\expandafter{\romannumeral4} in~\cite{yu2019lagrange}.

Finally, the master obtains the approximation results $\mathbf{Y}_i''=f'(\mathbf{X}_i'')\approx h''(\alpha_i'')$ for $i\in\{0,1,2\}$.

\textit{\textbf{Remark}~4}:~Compared to the LCC scheme, our BRI code has a lower computational load. In particular, the encoding and decoding functions of the BRI code involve fewer multiplication operations compared to the LCC scheme. Notably, the BRI code can recover the final result using any returned computational results. In contrast, the LCC scheme can only recover the final result when the number of received results from workers exceeds the required recovery threshold. Moreover, the LCC scheme, designed based on Lagrange interpolation, is only applicable to finite fields and cannot be applied to real number fields.

The encoding function of the LCC scheme without considering colluding workers is expressed as follows:
\begin{equation}
\begin{split}
h_{\rm{LCC}}(x)=\sum_{i\in\mathcal{M}}\mathbf{X}_i\prod_{k\in\mathcal{M}\backslash\{i\}}\frac{x-\alpha_k}{\alpha_i-\alpha_k}.
\label{eqDedfuncLCC}
\end{split}
\end{equation}
From Eq.~\eqref{eqDedfunc} and Eq.~\eqref{eqDedfuncLCC}, it is evident that the number of multiplication operations involved in the encoding function of the BRI code is significantly smaller than that required by the encoding function of the LCC scheme.

\subsubsection{\textbf{Example~\uppercase\expandafter{\romannumeral4}.3.}~(EP code~\cite{yu2020straggler})}
In this example, the master performs the same computing task, i.e., $f'(\mathbf{X})=\mathbf{X}^T\mathbf{X}$ using the EP code in a distributed computing system, where $\mathbf{X}=(\mathbf{X}_0, \mathbf{X}_1, \mathbf{X}_2)$. Thus, the objective of the master is to obtain $f'(\mathbf{X})=\mathbf{X}_i^T\mathbf{X}_i$ for $i\in\{0,1,2\}$.

Firstly, the encoding functions are given by
\begin{equation}
\begin{split}
\hbar_A(x)&=\mathbf{X}^T_0+\mathbf{X}^T_1x+\mathbf{X}^T_2x^2,\\
\hbar_B(x)&=\mathbf{X}_0+\mathbf{X}_1x^3+\mathbf{X}_2x^6.
\label{eqEdfuncExam3}
\end{split}
\end{equation}
Then, the master selects $9$ distinct elements ($x_0,x_1,\ldots,x_8$) from a finite field $\mathbb{F}_q$. Subsequently, the master encodes the input matrices $(\mathbf{X}_0, \mathbf{X}_1, \mathbf{X}_2)$ using the encoding functions given in Eq.~\eqref{eqEdfuncExam3}, and obtains $\hbar_A(x_i)$ and $\hbar_B(x_i)$ for $i\in\{0,1,\ldots,8\}$. After that, the master sends the encoded data $\hbar_A(x_i)$ and $\hbar_B(x_i)$ to worker $W_i$ for $i\in\{0,1,\ldots,8\}$.

Each worker $W_i$ receives the encoded data $\hbar_A(x_i)$ and $\hbar_B(x_i)$, and then performs the following computational task:
\begin{equation}
\begin{split}
\label{eqsubtask}
\tilde{\hbar}(x_i)=&\hbar_A(x_i)\hbar_B(x_i)\\
=&\mathbf{X}^T_0\mathbf{X}_0+\mathbf{X}^T_1\mathbf{X}_0x_i+\mathbf{X}^T_2\mathbf{X}_0x_i^2+\mathbf{X}^T_0\mathbf{X}_1x_i^3\\
&+\mathbf{X}^T_1\mathbf{X}_1x_i^4+\mathbf{X}^T_2\mathbf{X}_1x_i^5+\mathbf{X}^T_0\mathbf{X}_2x_i^6\\
&+\mathbf{X}^T_1\mathbf{X}_2x_i^7+\mathbf{X}^T_2\mathbf{X}_2x_i^8.
\end{split}
\end{equation}

Next, worker $W_i$ returns the result $\tilde{\hbar}(x_i)$ to the master. Note that Eq.~\eqref{eqsubtask} is a form of the $8$th-degree polynomial $\tilde{\hbar}(x)$ at point $x=x_i$, where $\tilde{\hbar}(x)$ is expressed as
\begin{equation}
\begin{split}
\label{eqsubtaskup}
\tilde{\hbar}(x)=&\mathbf{X}^T_0\mathbf{X}_0+\mathbf{X}^T_1\mathbf{X}_0x+\mathbf{X}^T_2\mathbf{X}_0x^2+\mathbf{X}^T_0\mathbf{X}_1x^3\\
&+\mathbf{X}^T_1\mathbf{X}_1x^4+\mathbf{X}^T_2\mathbf{X}_1x^5+\mathbf{X}^T_0\mathbf{X}_2x^6\\
&+\mathbf{X}^T_1\mathbf{X}_2x^7+\mathbf{X}^T_2\mathbf{X}_2x^8.
\end{split}
\end{equation}
From Eq.~\eqref{eqsubtaskup}, we find that the coefficients of powers of $x$ in $\tilde{\hbar}(x)$ include $\mathbf{X}^T_0\mathbf{X}_0$, $\mathbf{X}^T_1\mathbf{X}_1$, and $\mathbf{X}^T_2\mathbf{X}_2$, which are the desired computational results required by the master. The master is able to evaluate these coefficients of $\tilde{\hbar}(x)$ by polynomial interpolation over any $9$ computational results from workers. Therefore, the recovery threshold of the EP code is $9$ under this setting.

As discussed above, the recovery threshold achieved by the EP code is much larger than those of the BRI and LCC schemes. This inevitably increases the task completion latency in a CDC system. To successfully recover the final computation result, the LCC and the EP schemes are required to collect no fewer than $5$ and $9$ returned sub-results, respectively. In contrast to the LCC and the EP schemes, our BRI code is able to recover the final desired result utilizing any returned computational results. Evidently, our BRI code provides remarkable flexibility in result recovering. Furthermore, the decoding complexity of the BRI, LCC, and EP schemes is $\mathcal{O}((k-d)dst)$, $\mathcal{O}(st\log^2m\log\log m)$, and $\mathcal{O}(st\log^2m^2\log\log m^2)$~\cite{qiu2024secure,dutta2020optimal,qiu2024coded}, respectively. Consequently, the BRI code exhibits lower decoding complexity than the LCC and EP schemes. This confers a performance superiority on the BRI code.

\section{Application: The BRI code For Linear Regression}
In this section, we present a novel gradient coding algorithm for the linear regression problem, based on the proposed BRI code.

\subsection{Motivation}
In recent years, distributed machine learning has gained considerable attention. It partitions the model training task into multiple subtasks, distributes them to different computing nodes for parallel processing, and subsequently aggregates the results from all nodes to accelerate the overall training process. However, the presence of straggling computing nodes (stragglers), which complete their computing tasks much more slowly than normal nodes, significantly increases the training time.

Our proposed BRI code is able to efficiently address this problem due to its strong resilience against stragglers. By employing the BRI code, the final gradient can be approximately reconstructed using partially returned task results from computing nodes. Consequently, the model training time can be substantially reduced by eliminating the requirement to wait for all computing nodes to return their task results.

Given a large-scale dataset $\mathbf{A}\in\mathbb{R}^{s\times t}$ with corresponding labels $\mathbf{y}\in\mathbb{R}^{t}$, the linear regression problem aims to minimize the following loss function:
\begin{equation}
\label{eqLinear}
\min_{\mathbf{w}}\mathcal{L}(\mathbf{w})\triangleq\frac{1}{2}\min_{\mathbf{w}}\parallel\mathbf{Aw}-\mathbf{y}\parallel_{2}^{2},
\end{equation}
where $\mathbf{w}$ is the weight vector that needs to be solved.

Note that Eq.~\eqref{eqLinear} is convex with respect to $\mathbf{w}$. Thus, we can obtain the optimal the weight vector using the gradient descent algorithm. Specifically, the gradient of Eq.~\eqref{eqLinear} is $\triangledown\mathcal{L}(\mathbf{w})=\mathbf{A}^T(\mathbf{Aw-y})$. Then, we iteratively update the weight $\mathbf{w}$ using the following equation:
\begin{equation}
\begin{split}
\label{eqitera}
\mathbf{w}^{(t+1)}&=\mathbf{w}^{(t)}-\eta^{(t)}\triangledown\mathcal{L}(\mathbf{w}^{(t)})\\
&=\mathbf{w}^{(t)}-\eta^{(t)}\mathbf{A}^T(\mathbf{Aw}^{(t)}-\mathbf{y}),
\end{split}
\end{equation}
where $\eta^{(t)}$ is the learning rate in the $t$-th iteration.

In the next section, we propose a gradient coding algorithm that applies the BRI code to compute Eq.~\eqref{eqitera} in a distributed computing fashion.

\subsection{BRI-Based Algorithm for Linear Regression}
We now present an efficient BRI-based gradient coding algorithm for linear regression in a master-worker computing system. The system parameters are configured identically to those in Section~\uppercase\expandafter{\romannumeral3}-B.

Firstly, the master divides the large-scale input matrix $\mathbf{A}\in\mathbb{R}^{s\times t}$ into $m+1$ equal-sized sub-matrices such that $\mathbf{A}=[\mathbf{A}^T_0, \mathbf{A}^T_1, \ldots, \mathbf{A}^T_m]^T$, where $\mathbf{A}_i\in\mathbb{R}^{\frac{s}{m+1}\times t}$ for $i\in\mathcal{M}$. When $s$ is not a multiple of $m+1$, zero-padding may be applied to the last block. Then, the master encodes the sub-matrices $\{\mathbf{A}_i\}_{i=0}^{m}$ using the encoding function as follows:
\begin{equation}
\begin{split}
u(x)=\frac{\sum_{i=0}^{m-d}\psi_i(x)\rho_i(x)}{\sum_{j=0}^{m-d}\psi_j(x)},
\label{eqencfunc}
\end{split}
\end{equation}
where $\psi_i(x)$ and $\rho_i(x)$ are as follows:
\begin{equation}
\begin{split}
\psi_i(x)=\prod_{j=0}^{i-1}(x-\sigma_j)\prod_{k=i+d+1}^{m}(\sigma_k-x),
\label{eqEdff2}
\end{split}
\end{equation}
and
\begin{equation}
\begin{split}
\rho_i(x)=\sum_{k=i}^{i+d}\prod_{j=i,j\neq k}^{i+d}\frac{x-\sigma_j}{\sigma_k-\sigma_j}\mathbf{A}_k,
\label{eqEdff3}
\end{split}
\end{equation}
where $\sigma_0,\sigma_1,\ldots,\sigma_m$ are any $m+1$ distinct values from the real field $\mathbb{R}$.

Consequently, the master obtains the encoded matrices by computing $\tilde{\mathbf{A}}_i=u(\beta_i)$ for $i\in\mathcal{N}$, where $\beta_0,\beta_1,\ldots,\beta_{N-1}$ are $N$ distinct Chebeshev
points of the second kind from the real field $\mathbb{R}$. Using Chebyshev nodes of the second kind for interpolation can effectively avoid the Runge phenomenon and prevent severe oscillations near the endpoints~\cite{jahani2023berrut}. Moreover, the corresponding Lebesgue constant grows slowly, which ensures numerical stability in the interpolation process and enables the interpolating polynomial to smoothly approximate the target function~\cite{berrut2004barycentric,guttel2012convergence}. It is demonstrated in~\cite{guttel2012convergence} that Chebyshev points provide superior approximation performance. Note that $\{\sigma_i\}^{m}_{i=0}\cup\{\beta_i\}_{i=0}^{N-1}=\varnothing$. Then, the master sends the encoded matrix $\tilde{\mathbf{A}}_i$ to each worker $W_i$ for all $i\in\mathcal{N}$, and this process is performed only once. The encoded matrix $\tilde{\mathbf{A}}_i$ is stored locally at the corresponding worker $W_i$.

\begin{figure}[!t]
\centering
\includegraphics[width=3.3in]{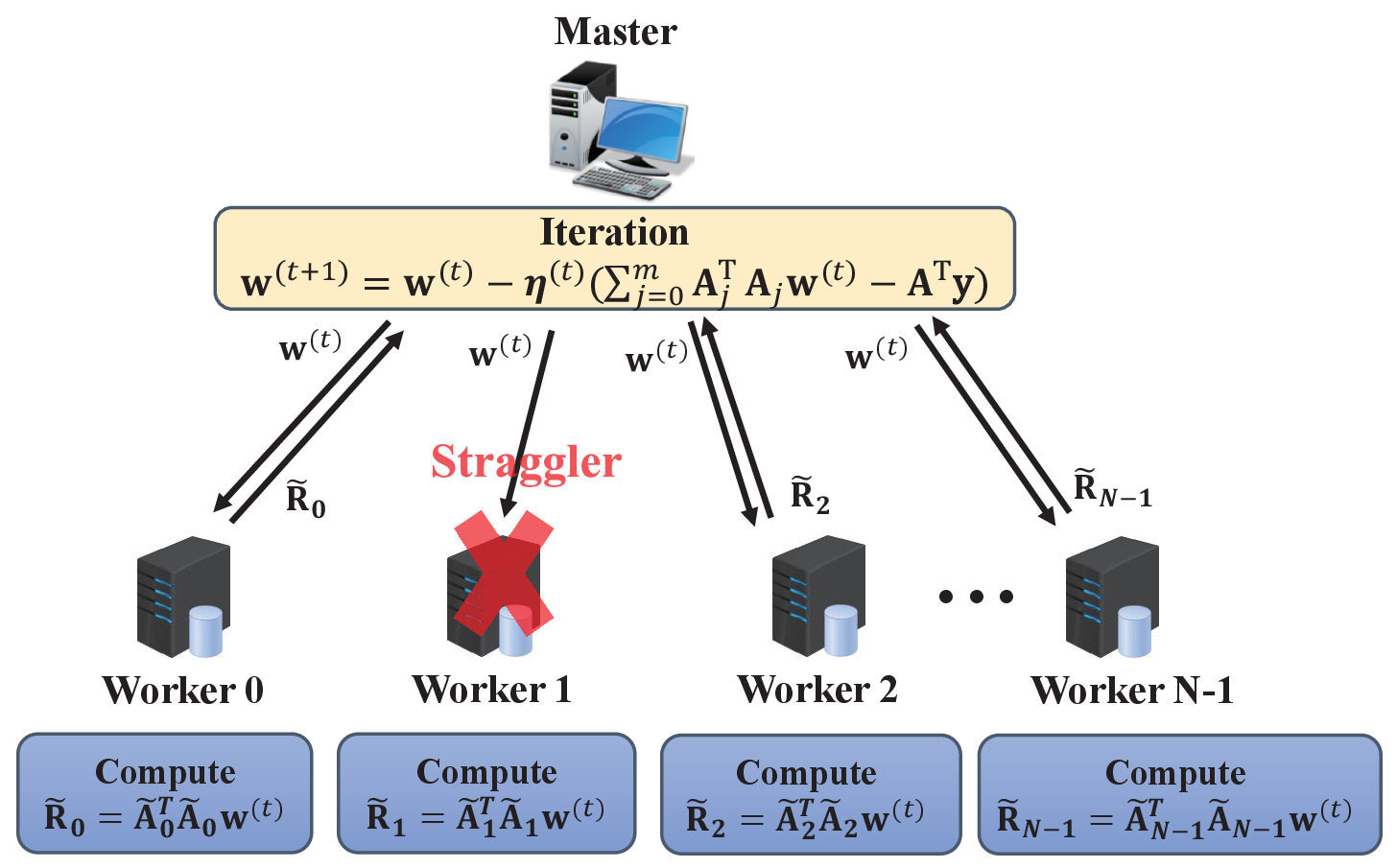}
\caption{An illustration of a BRI-based framework for linear regression.}
\label{fig:BRIforLR}
\end{figure}

As shown in Fig.~\ref{fig:BRIforLR}, the master sends the current weight $\mathbf{w}$ to each worker in each iteration.
Each worker $W_i$ receives the current weight parameter $\mathbf{w}$ and then executes its computing task, defined as $\Upsilon(\tilde{\mathbf{A}}_i)=\tilde{\mathbf{R}}_i=\tilde{\mathbf{A}}^T_i\tilde{\mathbf{A}}_i\mathbf{w}$. After completing the computing task, worker $W_i$ immediately returns the computed result $\tilde{\mathbf{R}}_i$ to the master. In the system, it is critical to note that not all workers return computed results to the master promptly. Some workers, termed stragglers, may either return computed results with substantial delays or fail to respond entirely.

The master waits and receives the returned computational result $\tilde{\mathbf{R}}_i$ from worker $W_i$ for $i\in\mathcal{P}$. $\mathcal{P}$ denotes the set of indices corresponding to the workers that successfully return their results to the master. Based on the returned results, we construct a decoding function $\Psi(x)$, as follows:
\begin{equation}
\begin{split}
\Psi(x)=\frac{\sum_{i=0}^{p-d}\psi'_i(x)\rho'_i(x)}{\sum_{j=0}^{p-d}\psi'_j(x)},
\label{eqDecodingFunc}
\end{split}
\end{equation}
$\psi'_i(x)$ and $\rho'_i(x)$ are as follows:
\begin{equation}
\begin{split}
\psi'_i(x)=\prod_{j=0}^{i-1}(x-\beta_{p_j})\prod_{\kappa=i+d+1}^{p}(\beta_{p_\kappa}-x),
\end{split}
\end{equation}
and
\begin{equation}
\begin{split}
\rho'_i(x)=\sum_{\kappa=i}^{i+d}\prod_{j=i,j\neq \kappa}^{i+d}\frac{x-\beta_{p_j}}{\beta_{p_\kappa}-\beta_{p_j}}\tilde{\mathbf{R}}_{p_\kappa},
\end{split}
\end{equation}
where $p=|\mathcal{P}|$ and $p_i$ is the $i$th element of the $\mathcal{P}$. It is observed that Eq.~\eqref{eqDecodingFunc} is designed to approximately interpolate $\Upsilon(u(x))$ using the interpolation points $(\beta_i,\Upsilon(u(\beta_i))$ for $i\in\mathcal{M}$.

Consequently, the master obtains the approximation results $\Psi(\sigma_i)\approx\Upsilon(\mathbf{A}_i)=\mathbf{A}^T_i\mathbf{A}_i\mathbf{w}$ for $i\in\mathcal{M}$ and computes $\mathbf{A}^T\mathbf{A}\mathbf{w}=\sum_{j=0}^{m}\mathbf{A}^T_i\mathbf{A}_i\mathbf{w}$. Using the same method, we obtain the value of $\mathbf{A}^T\mathbf{y}$. It should be noted that the value of $\mathbf{A}^T\mathbf{y}$ only needs to be computed once. Subsequently, the master updates the weight vector using Eq.~\eqref{eqitera}.

\section{Theoretical Analysis}
In this section, we provide a theoretical analysis of our BRI code, including its computational complexity and approximation error.

\subsection{Complexity Analysis}
\subsubsection{Encoding Complexity}
We compute the total encoding complexity by evaluating the function $r(x)$, given in Eq.~\eqref{eqEdfunc}. In Eq.~\eqref{eqEdfunc}, we observe that $r(x)$ is the sum of $(d+1)(m-d+1)$ matrices, each with dimension $s\times t$. Then, the encoding complexity for each worker is $\mathcal{O}\big((m-d)dst\big)$. Hence, the total encoding complexity for $N$ workers is $\mathcal{O}\big((m-d)dstN\big)$, where $N$ is the number of workers, $m$ is the number of submatrices, and $d$ is the parameter of the coding function~\eqref{eqEdfunc}.

\subsubsection{Decoding Complexity}
To compute the decoding complexity, we need to evaluate the function $h(x)$, given in Eq.~\eqref{eqDedfunc}. In Eq.~\eqref{eqDedfunc}, we observe that $h(x)$ is the sum of $(d+1)(k-d+1)$ matrices, each with dimension $s\times t$. Hence, the decoding complexity for the master is $\mathcal{O}\big((k-d)dst\big)$.

\subsubsection{Communication Complexity}
The communication complexity comprises two components: (i) master-to-worker communication and (ii) worker-to-master communication. For the first component, the master transmits $\mathcal{O}(st)$ symbols to each worker. Therefore, the total symbols of the master transmits to $N$ workers is $\mathcal{O}(stN)$. For the second component, worker $W_i$ that completes the assigned task $\tilde{\mathbf{Y}}_i=\tilde{\mathbf{X}}^T_i\tilde{\mathbf{X}}_i$ returns $\mathcal{O}(t^2)$ symbols to the master. Thus, the total number of symbols that the master receives from the fastest $k$ workers is $\mathcal{O}(kt^2)$.

\subsubsection{Each Worker's Computational Complexity}
Consider the computing task $\tilde{\mathbf{Y}}_i=\tilde{\mathbf{X}}^T_i\tilde{\mathbf{X}}_i$, where the dimension of $\tilde{\mathbf{X}}_i$ is $s\times t$. Therefore, the computational complexity for each worker is $\mathcal{O}(st^2)$.

\begin{figure}[!t]
    \centering
    \begin{minipage}[t]{0.24\textwidth}
        \centering
        \includegraphics[width=\linewidth]{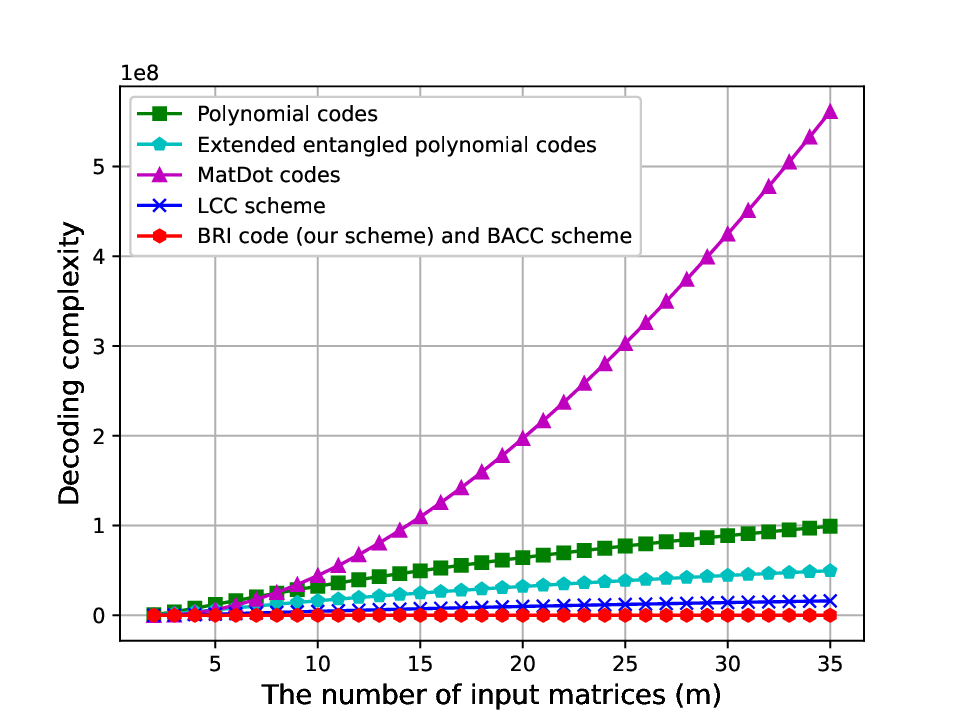}
        \caption{Comparison of the decoding complexity obtained by the BRI code, BACC scheme, LCC, MatDot, T-security EP and polynomial codes in a CDC system with parameters $s=1000$, and $m$ varying from $1$ to $36$.}
        \label{fig:decoding}
    \end{minipage}
    \hfill
    \begin{minipage}[t]{0.24\textwidth}
        \centering
        \includegraphics[width=\linewidth]{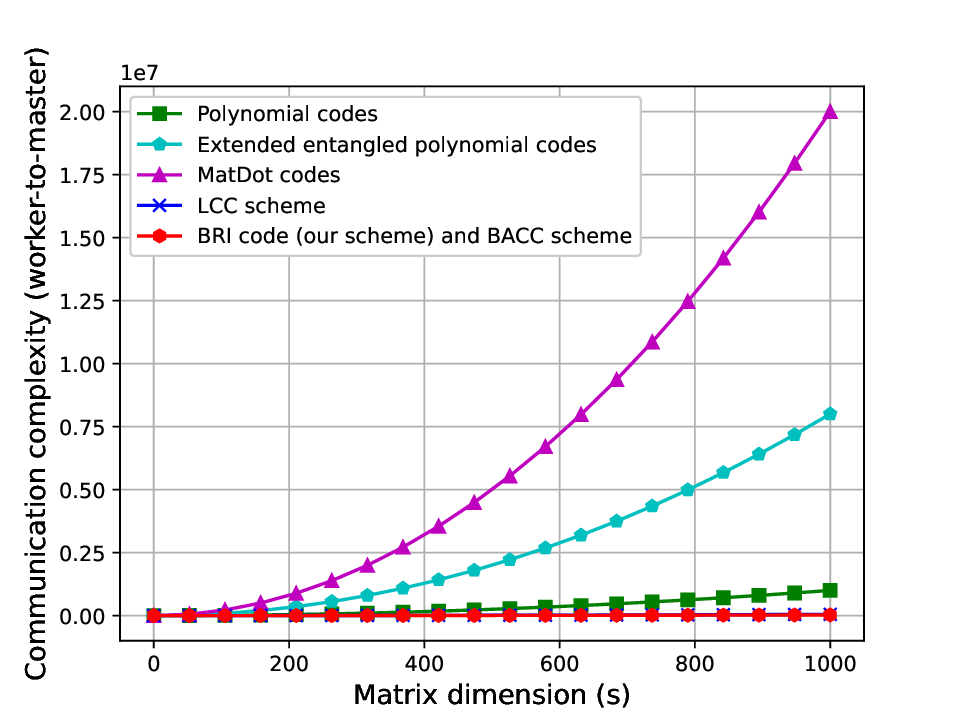}
        \caption{Comparison of the communication complexity of the BRI code, BACC scheme, LCC, MatDot, EEP and polynomial codes in a CDC system with parameters $k=10, m=20$ and $s$ varying from $1$ to $1000$.}
        \label{fig:communication}
    \end{minipage}
\end{figure}

As illustrated in Fig.~\ref{fig:decoding}, we compare the decoding complexity of our proposed BRI code with those of the BACC scheme, LCC, MatDot, extended entangled polynomial (EEP) codes~\cite{yu2020entangled} and polynomial codes, under parameters $s=1000$, and $m\in[1,36]$. Clearly, the decoding complexity of the BRI code is comparable to that of the BACC scheme, with both being lower than those of the other CDC schemes under the same parameter settings. This outcome is mainly stems from the similar forms of their decoding functions. Moreover, we observe that the decoding complexity of the polynomial codes is greater than that of the EEP codes but lower than that of the MatDot codes. The decoding complexity of the MatDot codes is the highest among all CDC schemes. In addition, the decoding complexity of the EEP codes is greater than that of the LCC scheme but lower than that of the polynomial codes. This outcome is mainly due to their recovery thresholds, matrix dimensions, and decoding functions~\cite{dutta2020optimal}. The decoding complexity increases as the recovery threshold or the matrix dimensions involved in the decoding function increase.

In Fig.~\ref{fig:communication}, we compare the communication complexity (worker-to-master) of our proposed BRI code with those of the BACC scheme, LCC, MatDot, and polynomial codes, under parameters $k=10, m=30$, and $s\in[1,1000]$. It is apparent that the BRI code achieves a communication complexity comparable to the BACC scheme, and both the schemes consistently outperform the other CDC schemes. In addition, the communication complexity of the MatDot codes is greater than those of the LCC scheme, polynomial codes, and EEP codes. The communication complexity of the MatDot codes is the highest among all CDC schemes. The primary reasons for this result are the recovery threshold and the matrix dimensions. The communication complexity increases as either the recovery threshold or the matrix dimensions increases.

\begin{figure*}
\centering
\subfigure[$n=10$]{
\label{fig5:a} 
\includegraphics[width=1.7in]{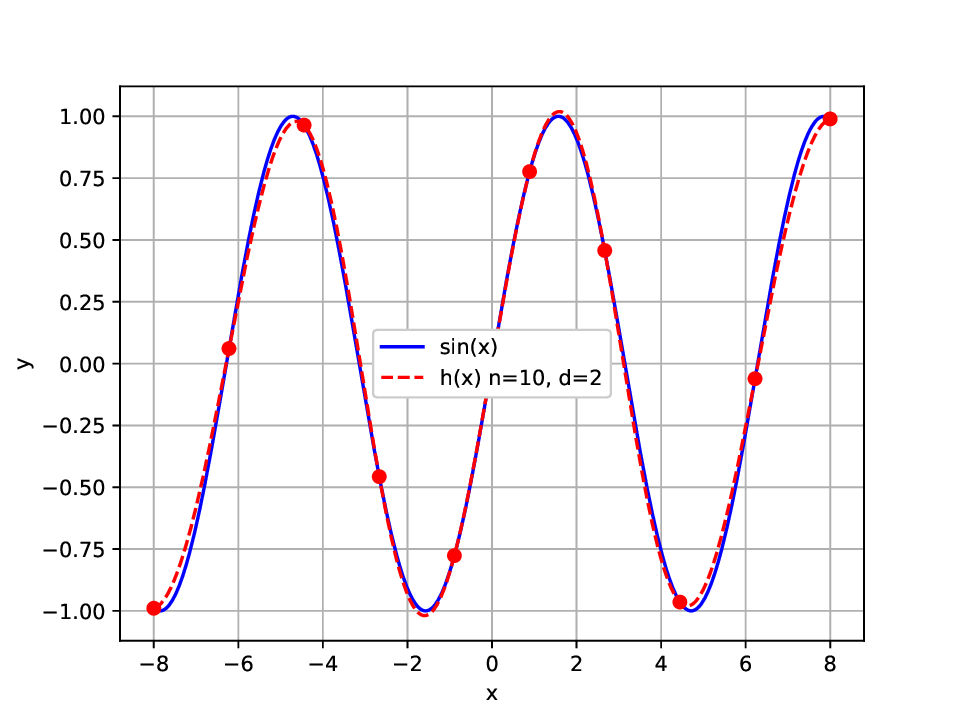}}
\subfigure[$n=15$]{
\label{fig5:b} 
\includegraphics[width=1.7in]{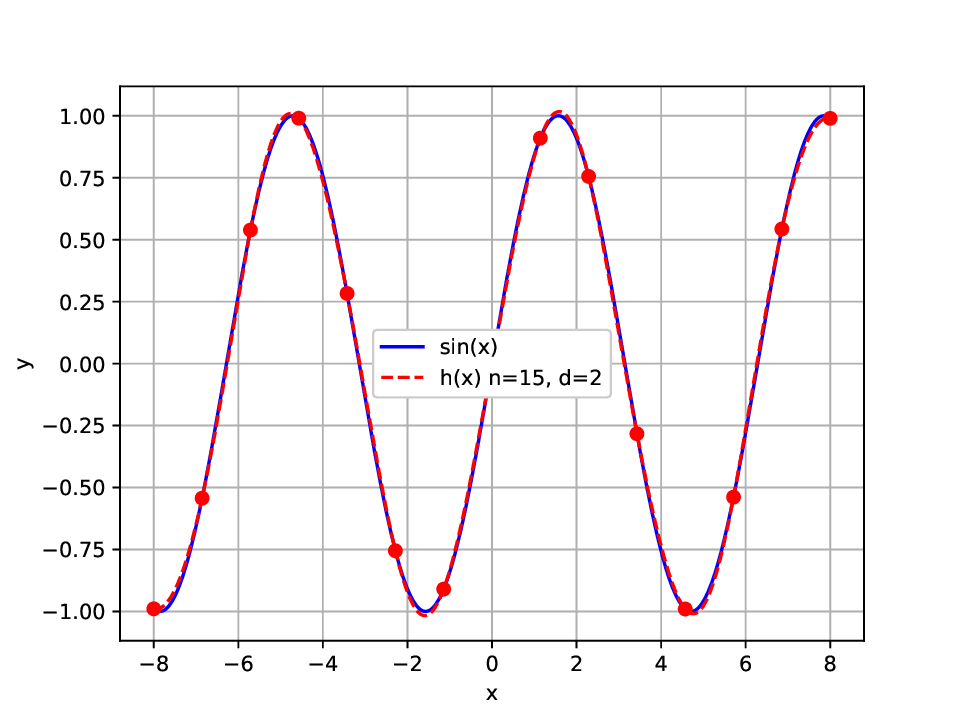}}
\subfigure[$n=20$]{
\label{fig5:c} 
\includegraphics[width=1.7in]{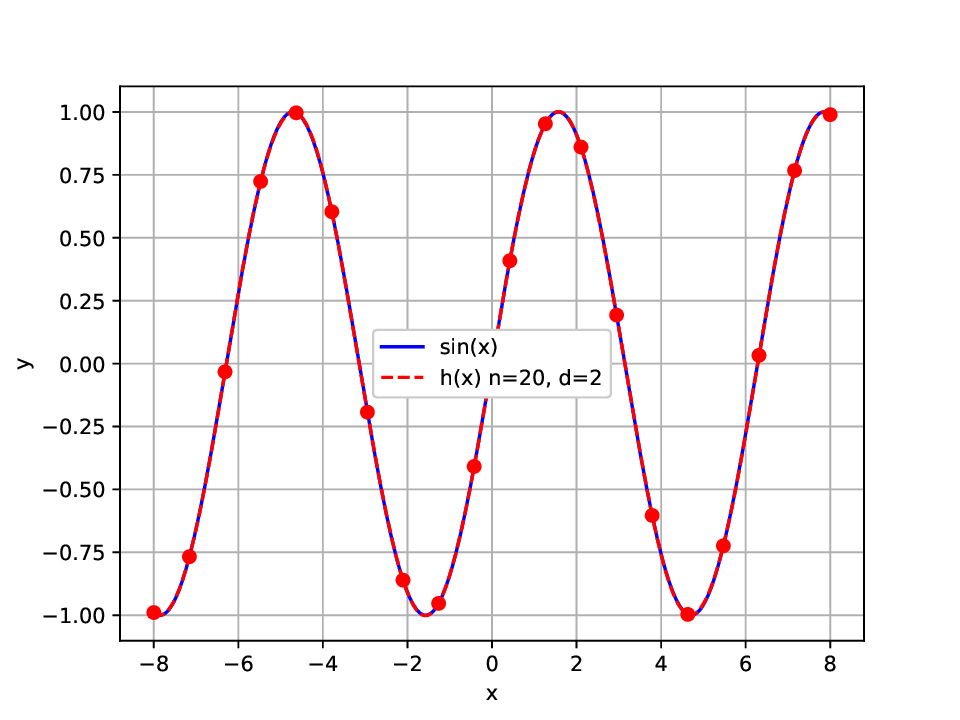}}
\subfigure[$n=25$]{
\label{fig5:d} 
\includegraphics[width=1.7in]{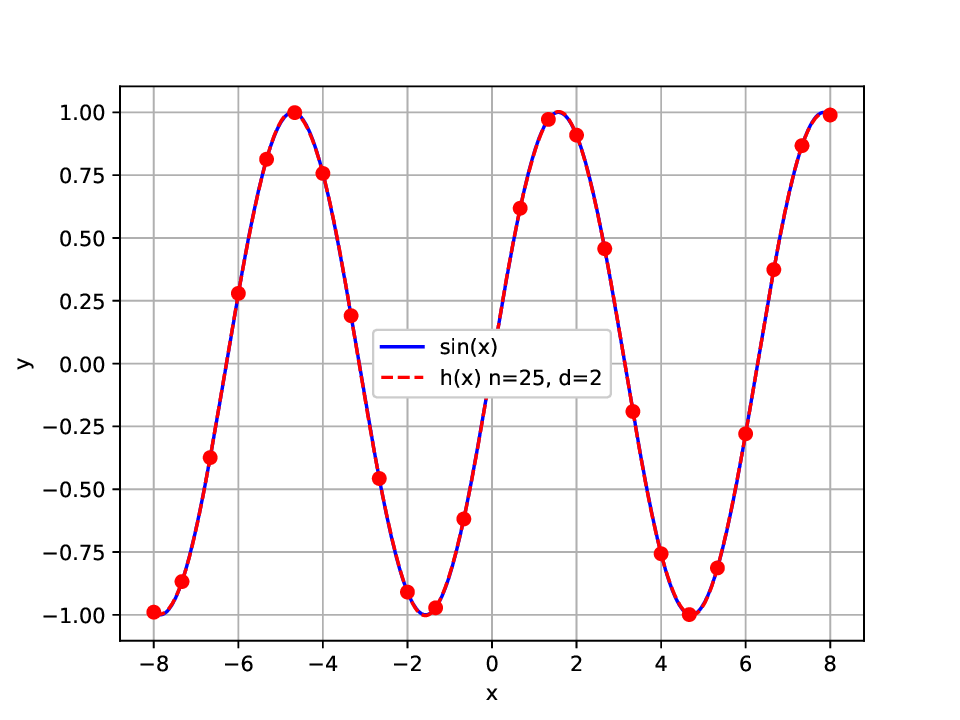}}
\caption{Comparison of interpolating the function $y=sin(x)$ using Eq.~\eqref{eqDedfunc} under the parameter $d=2$ and $n=10,15,20,$ and $25$.}
\label{fig2:interpolatSinFunc} 
\end{figure*}

\begin{table*}
\centering
\caption{COMPARISON OF MSE OBTAINED BY THE SCHEMES WITH $n=10, 15, 20$, AND $25$}
\label{Tab1}
\begin{tabular}{>{\centering\arraybackslash}p{20pt}
                >{\centering\arraybackslash}p{75pt}|
                >{\centering\arraybackslash}p{20pt}
                >{\centering\arraybackslash}p{75pt}|
                >{\centering\arraybackslash}p{20pt}
                >{\centering\arraybackslash}p{75pt}|
                >{\centering\arraybackslash}p{20pt}
                >{\centering\arraybackslash}p{75pt}}
\toprule
\multicolumn{2}{c|}{\( n = 10 \)} & \multicolumn{2}{c|}{\( n = 15 \)} & \multicolumn{2}{c|}{\( n = 20 \)} & \multicolumn{2}{c}{\( n = 25 \)}\\
\cmidrule(lr){1-2} \cmidrule(lr){3-4} \cmidrule(lr){5-6} \cmidrule(lr){7-8}
d & MSE & d & MSE & d & MSE & d & MSE\\
\midrule
0 & $2.639686\times 10^{-3}$ & 0 & $1.454047\times 10^{-3}$ & 0 & $7.957022\times 10^{-4}$ & 0 & $7.047475\times 10^{-4}$\\
1 & $4.260604\times 10^{-3}$ & 1 & $6.500152\times 10^{-4}$ & 1 & $1.405292\times 10^{-4}$ & 1 & $6.222679\times 10^{-5}$\\
1 & $1.374726\times 10^{-3}$ & 2 & $2.052476\times 10^{-4}$ & 2 & $3.304211\times 10^{-6}$ & 2 & $5.396126\times 10^{-6}$\\
3 & $8.152118\times 10^{-4}$ & 3 & $1.974928\times 10^{-5}$ & 3 & $5.091522\times 10^{-6}$ & 3 & $4.734734\times 10^{-7}$\\
4 & $6.261250\times 10^{-3}$ & 4 & $6.440430\times 10^{-5}$ & 4 & $7.288025\times 10^{-7}$ & 4 & $1.752035\times 10^{-7}$\\
5 & $7.207478\times 10^{-3}$ & 5 & $2.136655\times 10^{-5}$ & 5 & $2.653058\times 10^{-7}$ & 5 & $1.962853\times 10^{-9}$\\
6 & $3.085703\times 10^{-4}$ & 6 & $2.993368\times 10^{-6}$ & 6 & $2.653058\times 10^{-7}$ & 6 & $1.436792\times 10^{-8}$\\
7 & $1.252588\times 10^{-2}$ & 7 & $3.115002\times 10^{-5}$ & 7 & $4.220574\times 10^{-11}$ & 7 & $1.123866\times 10^{-9}$\\
8 & $5.237353\times 10^{-2}$ & 8 & $1.833997\times 10^{-5}$ & 8 & $6.094012\times 10^{-8}$ & 8 & $7.100354\times 10^{-10}$\\
9 & $5.237353\times 10^{-2}$ & 9 & $7.248981\times 10^{-7}$ & 9 & $1.280122\times 10^{-8}$ & 9 & $4.753900\times 10^{-10}$\\
\bottomrule
\end{tabular}
\end{table*}

\subsection{Approximation Error Analysis}
\begin{thm}
\label{thm0}
\!(\!\!\cite{floater2007barycentric})~Suppose that $d\geq 1$, the barycentric rational interpolant $h(x)$ in Eq.~\eqref{eqDedfunc} is used as the interpolation formula for a continuous function $f\in\mathbf{\mathcal{C}}^{d+2}[a,b]$ with a $(d+2)$ derivative. If $n-d$ is odd, then we obtain
\begin{equation}
\parallel h(x)-f(x)\parallel\leq \lambda^{d+1}(b-a)\frac{\parallel f^{(d+2)}(x)\parallel}{d+2}.
\end{equation}
If $n-d$ is even, then
\begin{equation}
\begin{split}
&\parallel h(x)-f(x)\parallel\\
&\leq\lambda^{d+1}\bigg[(b-a)\frac{\parallel f^{(d+2)}(x)\parallel}{d+2}+\frac{\parallel f^{(d+1)}(x)\parallel}{d+1}\bigg],
\end{split}
\end{equation}
where $\lambda\triangleq\max_{0\leq i\leq n-1}(x_{i+1}-x_i)$ and $n$ represents the number of interpolation points.
\end{thm}
\begin{IEEEproof}
The proof of Theorem~\ref{thm0} is provided in~\cite{floater2007barycentric}.
\end{IEEEproof}

\begin{rem}
A larger parameter $d$ corresponds to using higher-degree local polynomial interpolation, which enhances approximation accuracy by more effectively capturing local variations of the target function. On the other hand, increasing the number of interpolation points $n$ leads to denser sampling, thereby reducing the overall error, as shorter intervals between points improve the precision of local approximations. The approximation accuracy is jointly determined by $d$ and $n$.
\end{rem}

\begin{thm}
\label{thm1}
Given a CDC system consisting of a master, $N$ workers and $S$~$(S<N-2)$ stragglers, the decoding function $\Psi(x)$ in Eq.~\eqref{eqDecodingFunc} and a continuous $(d+2)$ derivative function $\varpi(x)=\Upsilon(u(x))\in\mathbf{\mathcal{C}}[-\frac{d+2}{2},\frac{d+2}{2}]$, the approximation error of the interpolation using the BRI-based linear regression algorithm is upper bounded as
\begin{equation}
\parallel \Psi(x)-\varpi(x)\parallel\leq \sin(\frac{(S+1)\pi}{2N})^{d+1}\parallel \varpi^{(d+2)}(x)\parallel,
\end{equation}
if $N-S$ is odd, and
\begin{equation}
\begin{split}
&\parallel\Psi(x)-\varpi(x)\parallel\\
&\leq\sin(\frac{(S+1)\pi}{2N})^{d+1}\bigg[\parallel \varpi^{(d+2)}(x)\parallel+\frac{\parallel \varpi^{(d+1)}(x)\parallel}{d+1}\bigg],\quad~~
\end{split}
\end{equation}
if $N-S$ is even.
\end{thm}
\begin{IEEEproof}
From Theorem~1, we have
\begin{equation}
\parallel \Psi(x)-\varpi(x)\parallel\leq \lambda^{d+1}\parallel \varpi^{(d+2)}(x)\parallel,
\label{eq33}
\end{equation}
if $N-S$ is odd, and
\begin{equation}
\begin{split}
&\parallel \Psi(x)-\varpi(x)\parallel\leq\lambda^{d+1}[\parallel \varpi^{(d+2)}(x)\parallel+\frac{\parallel \varpi^{(d+1)}(x)\parallel}{d+1}],
\label{eq34}
\end{split}
\end{equation}
if $N-S$ is even and where $\lambda\triangleq\max_{0\leq k\leq n-1}(x_{k+1}-x_k)$.

In Section~\uppercase\expandafter{\romannumeral5}-B, the master collects $p$ interpolation points and sorts them into a ordered set $\mathcal{X}=\{x_k\}_{k=0}^{p-1}$. Note that $\mathcal{X}$ is a subset of the Chebyshev points of second kind, defined as $\tilde{\mathcal{X}}=\{\tilde{x}_i\}_{i=0}^{N-1}$. Thus, we have $\mathcal{X}\subset\tilde{\mathcal{X}}$ and
\begin{equation}
x_k=\tilde{x}_{i_k}=-\cos(\frac{i_k\pi}{N}),
\label{eqapp}
\end{equation}
where $i_k\geq k$.

Let $\ell(k)=x_{k+1}-x_k$. There exists a constant $\vartheta$ within the interval $[1,S+1]$ such that
\begin{equation}
\ell(k)=-\cos(\frac{(i_k+\vartheta)\pi}{N})+\cos(\frac{i_k\pi}{N}).
\label{equ001}
\end{equation}

The function $\ell(k)$ reaches its maximal value at $\frac{i_k\pi}{N}=\frac{\pi}{2}-\frac{\vartheta\pi}{2N}$. Thus, we can obtain
\begin{equation}
\ell=\max_{0\leq k\leq p-1}(x_{k+1}-x_k)=2\sin(\frac{\vartheta\pi}{2N})\leq\sin(\frac{(S+1)\pi}{2N}),
\label{eq002}
\end{equation}
due to the monotonic increasing nature of $\sin(x)$ on $[0,\pi/2]$.

Therefore, from Eqs.~\eqref{eq33},~\eqref{eq34}, and~\eqref{eq002}, we can obtain
\begin{equation}
\parallel \Psi(x)-\varpi(x)\parallel\leq \sin(\frac{(S+1)\pi}{2N})^{d+1}\parallel \varpi^{(d+2)}(x)\parallel,
\end{equation}
if $N-S$ is odd, and
\begin{equation*}
\begin{split}
&\parallel \Psi(x)-\varpi(x)\parallel\\
&\leq\sin(\frac{(S+1)\pi}{2N})^{d+1}\bigg[\parallel \varpi^{(d+2)}(x)\parallel+\frac{\parallel \varpi^{(d+1)}(x)\parallel}{d+1}\bigg],
\end{split}
\end{equation*}
if $N-S$ is even. This completes the proof.
\end{IEEEproof}

\section{Experiments}
In this section, we perform comprehensive experiments to evaluate the performance of the BRI code. Firstly, we discuss the approximation properties of the decoding function $h(x)$. Then, we investigate the effectiveness of the BRI code on a large-scale computational task over a collaborative MEC system.

\begin{figure*}
\centering
\subfigure[Scenario 1 ($S=3$)]{
\label{figg:a} 
\includegraphics[width=1.7in]{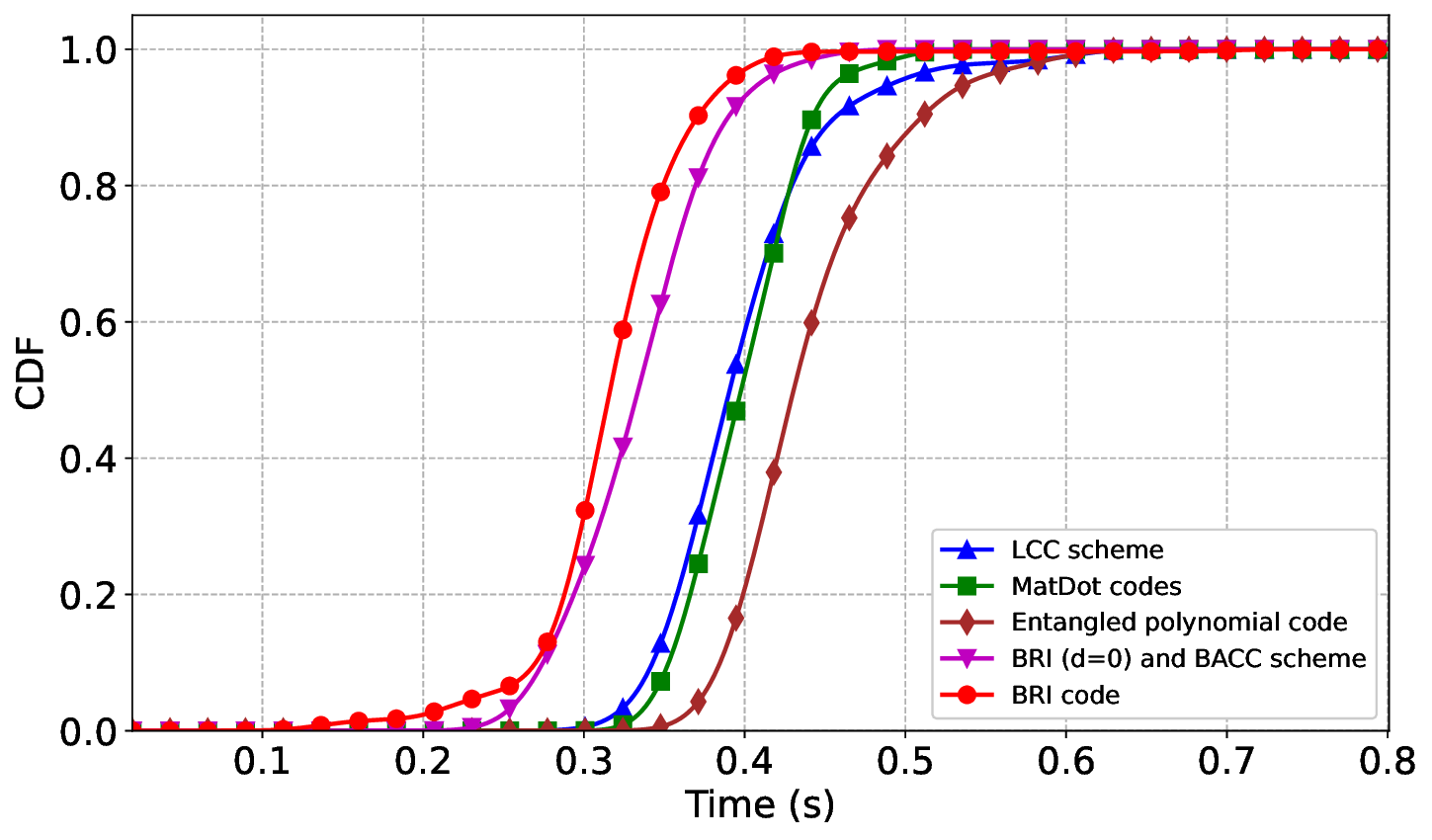}}
\subfigure[Scenario 2 ($S=5$)]{
\label{figg:b} 
\includegraphics[width=1.7in]{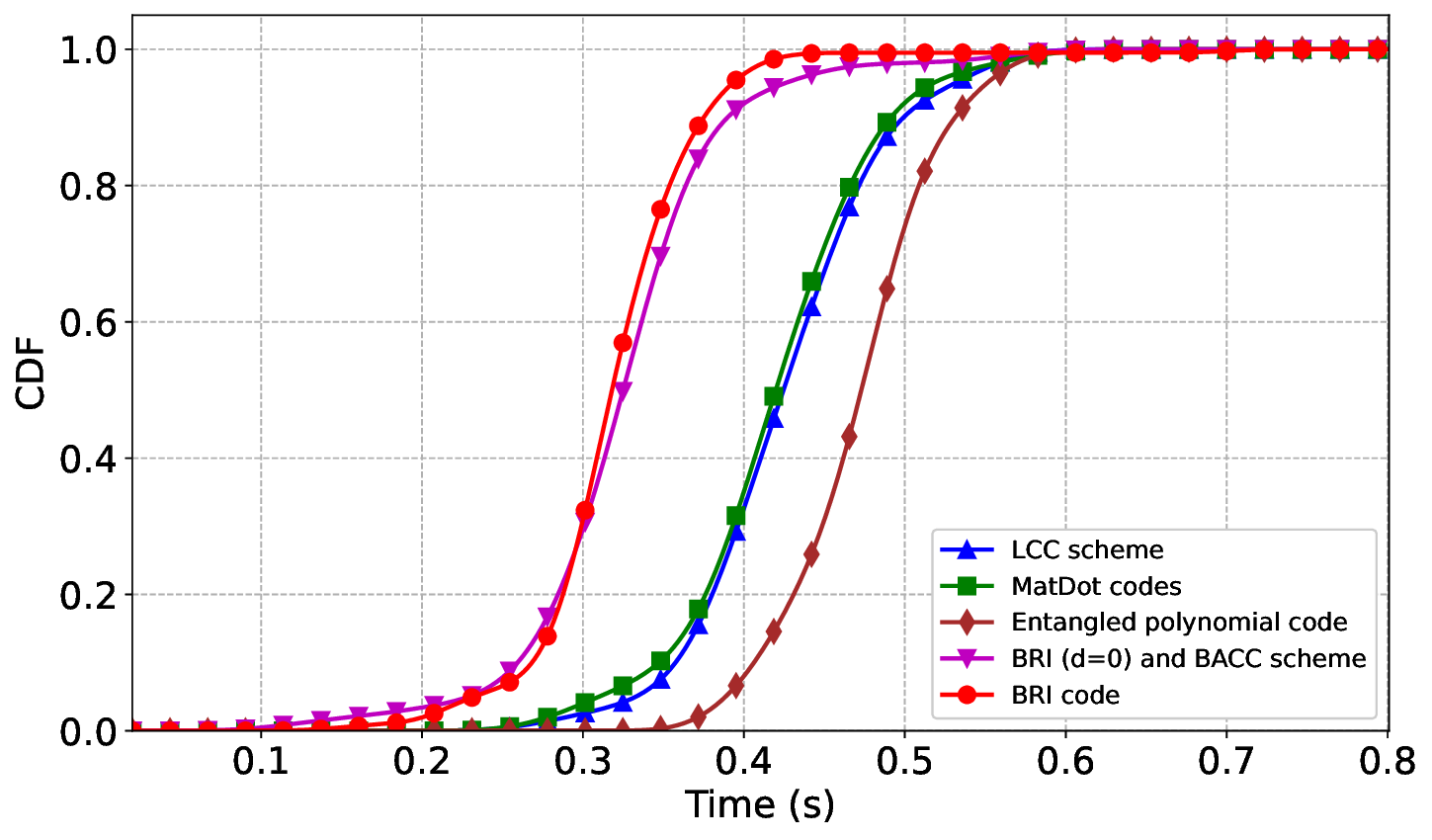}}
\subfigure[Scenario 3 ($S=7$)]{
\label{figg:c} 
\includegraphics[width=1.7in]{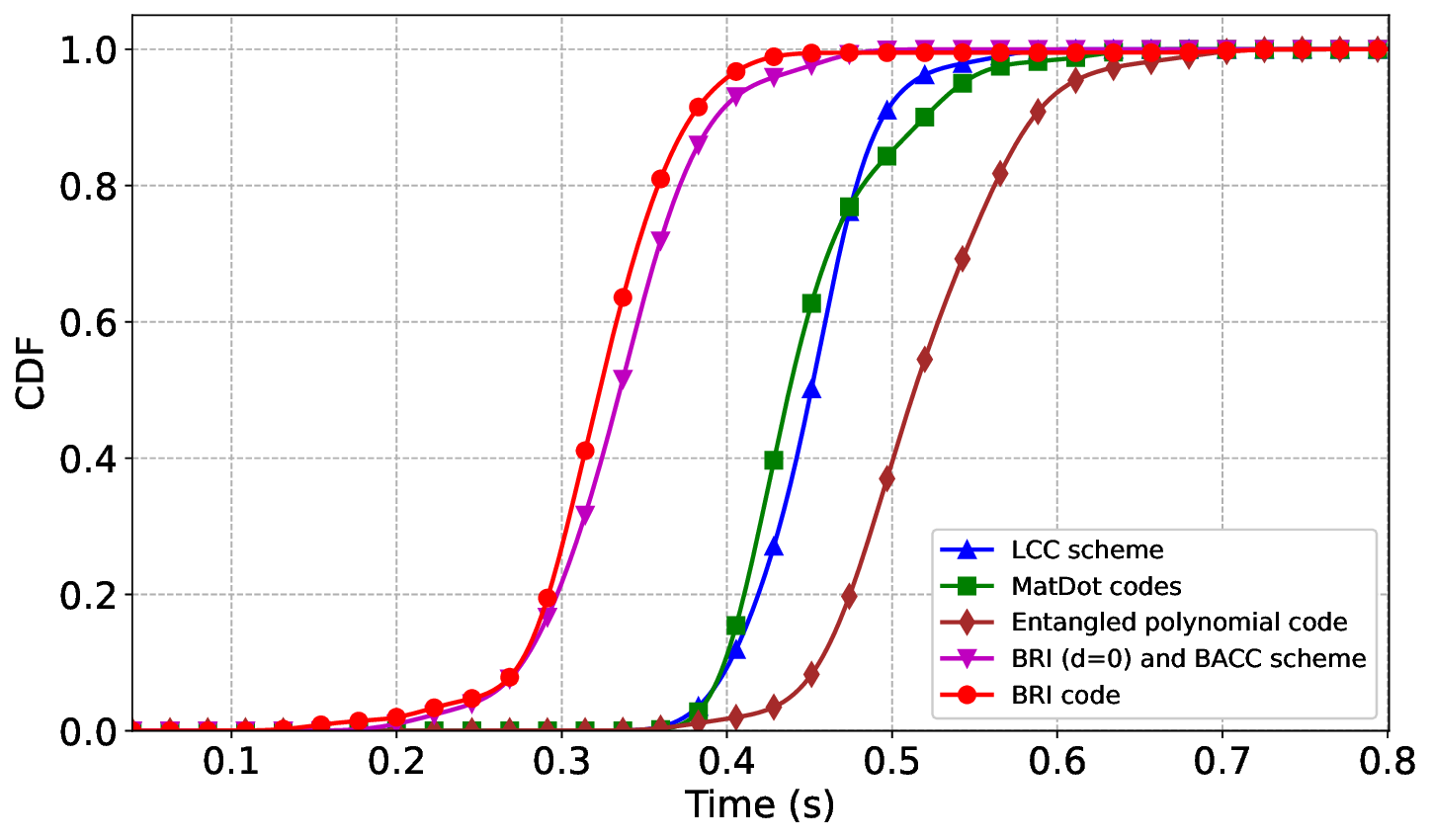}}
\subfigure[Scenario 4 ($S=9$)]{
\label{figg:d} 
\includegraphics[width=1.7in]{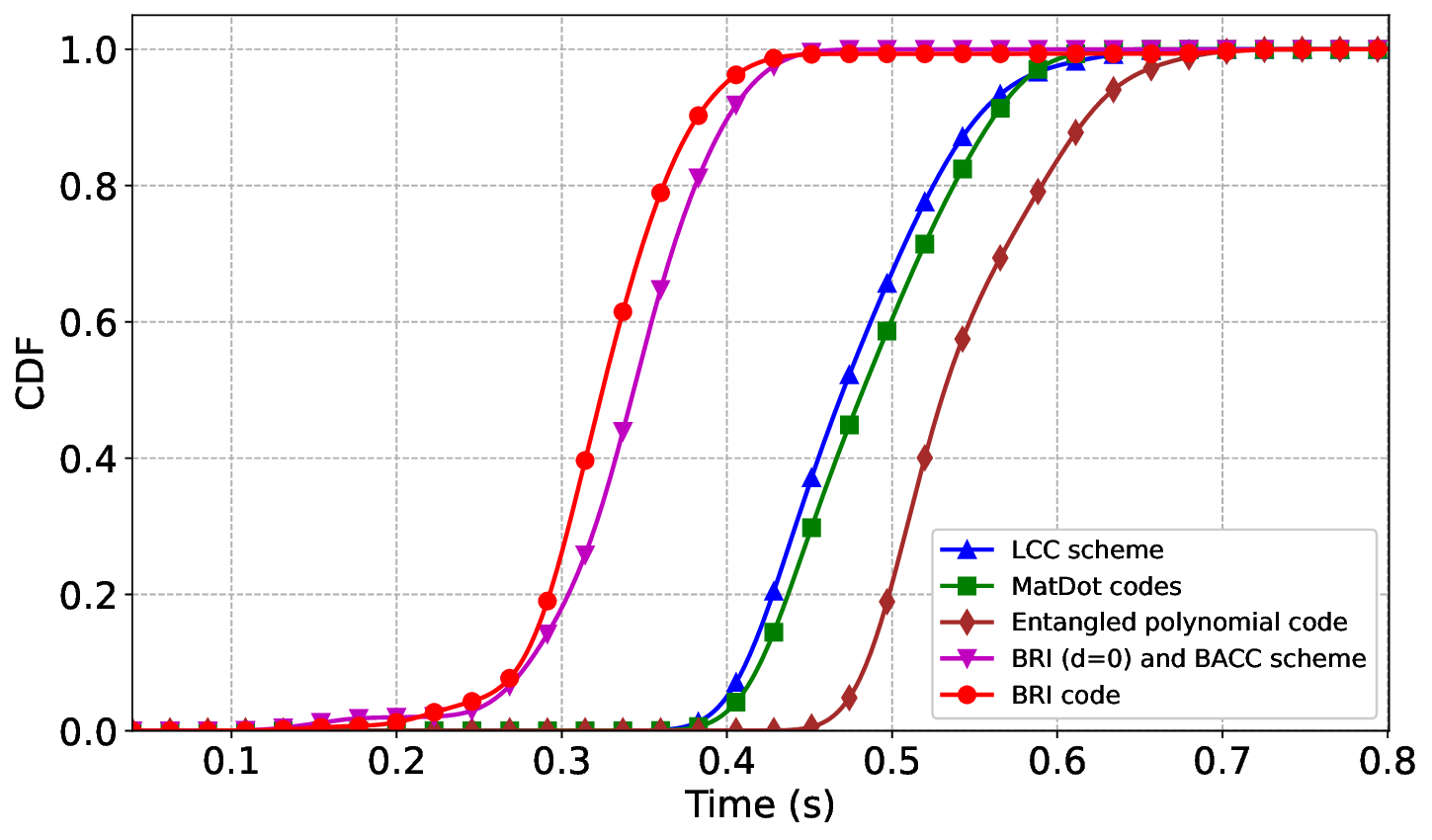}}
\caption{Comparison of CDFs of execution times achieved by the BACC scheme, LCC, MatDot, BRI, and EP codes, in CDC systems with the system parameters as $N=20$, and $S=3,5,7,9$.}
\label{figg:AverageTime}
\end{figure*}

\begin{table*}
\centering
\caption{Performance improvement in terms of $CDF=0.8$ of the proposed BRI compared with $S=3,5,7$, and $9$}
\label{Tab2}
\begin{tabular}{p{35pt}p{65pt}|p{35pt}p{65pt}|p{35pt}p{65pt}|p{35pt}p{65pt}}
\toprule
\multicolumn{2}{c|}{$S = 3$} & \multicolumn{2}{c|}{$S = 5$} & \multicolumn{2}{c|}{$S = 7$} & \multicolumn{2}{c}{$S = 9$}\\
\cmidrule(lr){1-2} \cmidrule(lr){3-4} \cmidrule(lr){5-6} \cmidrule(lr){7-8}
Baseline schemes & Relative Percentage & Baseline schemes & Relative Percentage & Baseline schemes & Relative Percentage & Baseline schemes & Relative Percentage\\
\midrule
BACC      & $5.50\%$ lower  & BACC      & $2.58\%$ lower  & BACC      & $3.68\%$ lower  & BACC      & $4.10\%$ lower\\
LCC       & $18.75\%$ lower & LCC       & $24.85\%$ lower & LCC       & $25.14\%$ lower & LCC       & $31.04\%$ lower\\
MatDot    & $18.60\%$ lower & MatDot    & $23.97\%$ lower & MatDot    & $25.73\%$ lower & MatDot    & $32.60\%$ lower\\
Entangled & $26.63\%$ lower & Entangled & $30.39\%$ lower & Entangled & $36.29\%$ lower & Entangled & $38.69\%$ lower\\
\bottomrule
\end{tabular}
\end{table*}

\begin{table*}
\centering
\caption{Performance improvement in terms of $CDF=1.0$ of the proposed BRI compared with the baseline $S=3,5,7$, and $9$}
\label{Tab3}
\begin{tabular}{p{35pt}p{65pt}|p{35pt}p{65pt}|p{35pt}p{65pt}|p{35pt}p{65pt}}
\toprule
\multicolumn{2}{c|}{$S = 3$} & \multicolumn{2}{c|}{$S = 5$} & \multicolumn{2}{c|}{$S = 7$} & \multicolumn{2}{c}{$S = 9$}\\
\cmidrule(lr){1-2} \cmidrule(lr){3-4} \cmidrule(lr){5-6} \cmidrule(lr){7-8}
Baseline schemes & Relative Percentage & Baseline schemes & Relative Percentage & Baseline schemes & Relative Percentage & Baseline schemes & Relative Percentage\\
\midrule
BACC      & $6.61\%$ lower  & BACC      & $7.73\%$ lower  & BACC      & $8.42\%$ lower  & BACC      & $7.21\%$ lower\\
LCC       & $22.97\%$ lower & LCC       & $25.29\%$ lower & LCC       & $29.94\%$ lower & LCC       & $33.64\%$ lower\\
MatDot    & $21.08\%$ lower & MatDot    & $26.30\%$ lower & MatDot    & $30.23\%$ lower & MatDot    & $32.03\%$ lower\\
Entangled & $25.31\%$ lower & Entangled & $26.50\%$ lower & Entangled & $36.71\%$ lower & Entangled & $39.14\%$ lower\\
\bottomrule
\end{tabular}
\end{table*}

\begin{figure}[!t]
\centering
\includegraphics[width=3.3in]{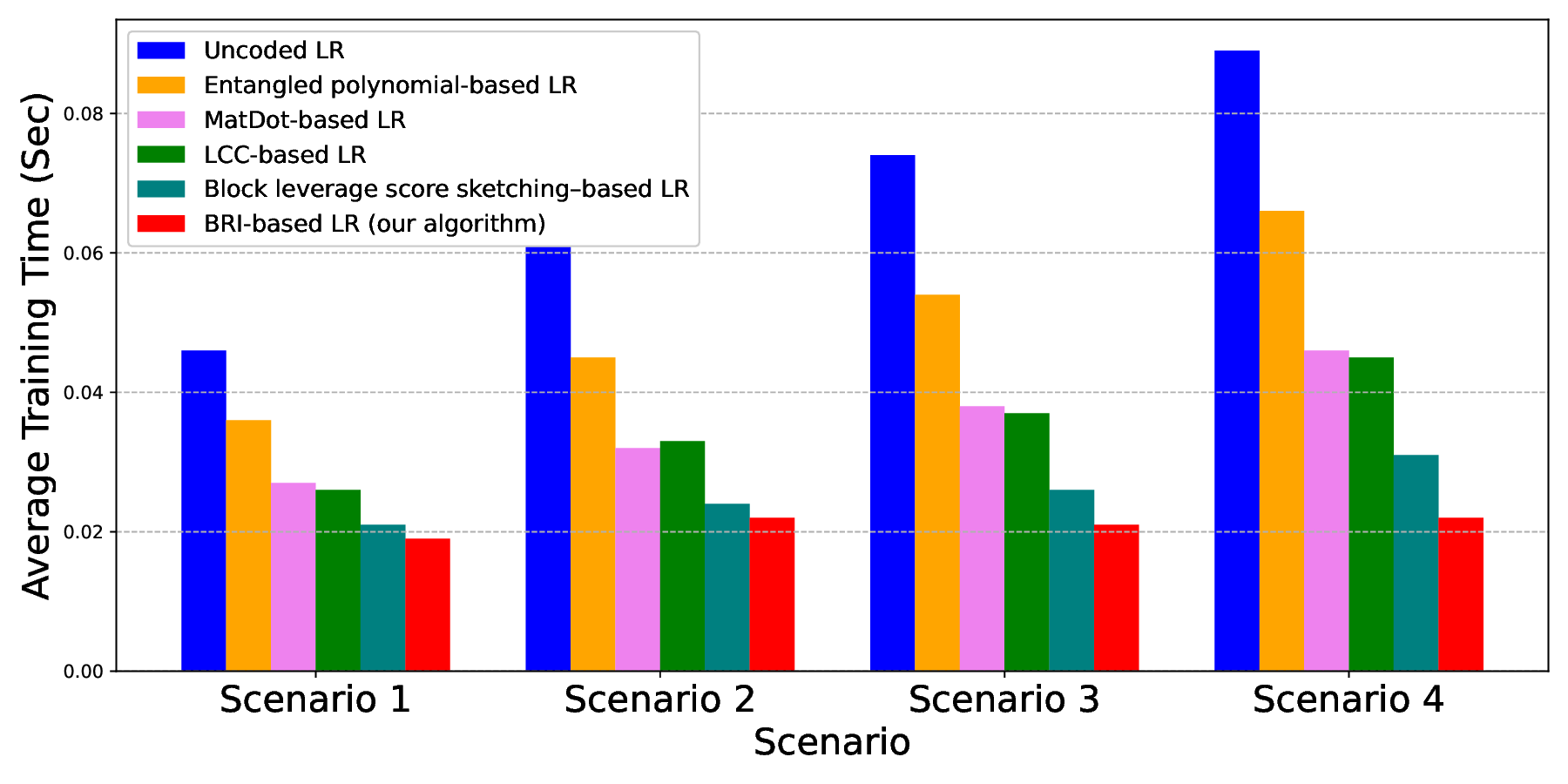}
\caption{Comparison of the average training time achieved by uncoded LR, EP-based LR, MatDot-based LR, LCC-based LR, block leverage score sketching-based LR and BRI-based LR algorithms, in different four scenarios, where scenario 1-4 correspond to $N=20, m=9$ with $S=3,5,7,$ and $9$, respectively.}
\label{fig:averagetime}
\end{figure}

\subsection{Experimental Settings}
Now, we proceed to reveal the advantages of the BRI code in performing a distributed computing task.

We consider a collaborative CDC system comprising a master and $N=20$ workers based on the MPI4PY package~\cite{rogowski2022mpi4py}. The experimental platform is equipped with an NVIDIA RTX 1050Ti, an Intel i5-7300HQ processor and 256GB DDR4 RAM. To simulate the behavior of stragglers, we utilize the \textit{sleep()} function from Python \textit{time} module. Moreover, we utilize the random functions (e.g., \textit{random.random()}) to generate the length, occurrence time, and location of the delay of the stragglers. The random function selects stragglers based on a uniform distribution. This approach provides a simplified approximation of the occurrence patterns of stragglers in real systems.

In the system, the master aims to approximately evaluate a computational task, i.e., $\mathbf{Y}=\mathbf{X}^T\mathbf{X}$ over a dataset  $\mathbf{X}=(\mathbf{X}_0, \mathbf{X}_1, \ldots, \mathbf{X}_9)$, where each element $\mathbf{X}_i$ for $i\in\{0,1,\ldots, 9\}$ has a dimension of $100\times 100$. To evaluate the proposed BRI code, we introduce the EP code~\cite{yu2020straggler}, MatDot codes~\cite{dutta2020optimal}, LCC scheme~\cite{yu2019lagrange}, and BACC scheme~\cite{jahani2023berrut}) as baseline schemes. We consider four scenarios as follows:
\begin{itemize}
\item [$\bullet$]
\textbf{Scenario 1:}~$N=20$, $m=9$ and $S=3$.
\item [$\bullet$]
\textbf{Scenario 2:}~$N=20$, $m=9$ and $S=5$.
\item [$\bullet$]
\textbf{Scenario 3:}~$N=20$, $m=9$ and $S=7$.
\item [$\bullet$]
\textbf{Scenario 4:}~$N=20$, $m=9$ and $S=9$.
\end{itemize}

In our experiments, the master is required to execute the task $50$ times in each scenario.

\subsection{Approximation Error Analysis}
In this subsection, we analyze the approximation error of the function $h(x)$ when interpolating the function $g(x)=sin(x)$ over values of $x$.

As shown in Fig.~\ref{fig2:interpolatSinFunc}, we employ the designed function $h(x)$ to interpolate the function $g(x)=sin(x)$ for $x\in[-8,8]$ based on parameters $d=2$ and $n\in\{10,15,20,25\}$. From Fig.~\ref{fig2:interpolatSinFunc}, we observe that our function $h(x)$ has a good approximation to the function $g(x)$ when $n\geq 10$. As the number of $n$ increases, the computed error tend to zero. Specifically, for $n\geq 10$, the interpolated function $h(x)$ closely approximates the original function $g(x)$. Therefore, we can achieve very small approximation error using only a few interpolation points by adjusting parameters $d$ and $n$.

\begin{table}
\centering
\begin{minipage}{0.45\textwidth}
\centering
\caption{Comparison of Computed Errors with \( n = 8 \)}
\label{Tabb8}
\begin{tabular}{p{40pt}p{40pt}p{100pt}}
\toprule
\( n \) & \( d \) & error \\
\midrule
8 &0 &$4.439392725\times 10^{-5}$\\
8 &1 &$4.614209911\times 10^{-6}$\\
8 &2 &$1.582230517\times 10^{-6}$\\
8 &3 &$2.679536593\times 10^{-7}$\\
8 &4 &$3.688160888\times 10^{-8}$\\
8 &5 &$8.881784197\times 10^{-8}$\\
8 &6 &$6.994405055\times 10^{-8}$\\
8 &7 &$1.554312234\times 10^{-9}$\\
8 &8 &$1.878497358\times 10^{-9}$\\
\bottomrule
\end{tabular}
\end{minipage}
\end{table}

To further examine the relationship between the approximation error and parameters $d$ and $n$, we apply the proposed interpolation method to the function $g(x)$ under parameters $n=10,15,20,25$ and $d=0,1,2,3,4,5,6,7,8,9$. As illustrated in Table~\ref{Tab1}, we compare the mean square error (MSE) achieved by the proposed interpolation method under the given parameter settings of $d$ and $n$. Furthermore, the minimum MSE can be achieved by adjusting parameters $d$ and $n$ while satisfying $0\leq d\leq n$. In Table~\ref{Tab1}, we observe that the MSE generally decreases with increasing $d$. However, at $d=7$, the MSE exceeds that at $d=0$, indicating that a larger $d$ does not guarantee a smaller MSE. Thus, the optimal $d$ should be selected based on $n$.

In conclusion, our experimental results demonstrate that our decoding function $h(x)$ has a superior performance in performing approximate decoding under a distributed computing environment. In particular, it has a high flexibility, enabling adjustable approximation accuracy according to parameters $d$ and $n$.


\subsection{Task Completion Time Analysis}
As illustrated in Fig.~\ref{figg:AverageTime}, we compare cumulative distribution functions (CDF) of waiting times for task completion using the proposed BRI code against various baseline schemes.

In Fig.~\ref{figg:AverageTime}, the CDF curve of our BRI code exhibits the fastest growth rate among all schemes, indicating significantly lower latency in computational task completion compared to the LCC, MatDot, EP, and BACC schemes under parameters $S=3,5,7$, and $9$. The BACC scheme exhibits a slightly slower curve growth compared to our BRI code. In fact, the BACC scheme is a special case of the BRI code. The growth rate of the EP scheme is the slowest among all schemes. Actually, the CDF of the waiting time is primarily influenced by the recovery threshold. The recovery threshold of the EP code is the highest. Therefore, the master must collect more computational results from the workers compared to other coding schemes. This significantly increases the waiting time for the EP scheme. On the contrary, the recovery threshold of the proposed BRI code is not strictly constrained. Thus, we can recover the final result using any received results from workers based on the BRI code. This significantly reduce the waiting time for the BRI code. Moreover, both the LCC scheme and MatDot scheme exhibit similar growth trends, positioned between the CDF curves of the BACC scheme and EP scheme.

Next, we conduct a comparative analysis of the efficacy of the BRI code, BACC scheme, LCC, MatDot, and EP codes in terms of CDF.
As shown in Table~\ref{Tab2}, when $S=3$, the proposed BRI code reaches a CDF value of $0.8$, which outperforms the BACC, LCC, MatDot, and EP schemes by $5.50\%$, $18.75\%$, $18.60\%$, and $26.63\%$, respectively. When $S=5$, the proposed BRI code reaches a CDF value of $0.8$, which outperforms the BACC, LCC, MatDot, and EP schemes by $2.58\%$, $24.85\%$, $23.97\%$, and $30.39\%$, respectively. When $S=7$, the proposed BRI code reaches a CDF value of $0.8$, which outperforms the BACC, LCC, MatDot, and EP schemes by $3.68\%$, $25.14\%$, $25.73\%$, and $36.29\%$, respectively. When $S=9$, the proposed BRI code reaches a CDF value of $0.8$, which outperforms the BACC, LCC, MatDot, and EP schemes by $4.10\%$, $31.04\%$, $32.60\%$, and $38.69\%$, respectively. In addition, the performance comparison of the BRI, BACC, LCC, MatDot, and EP schemes under parameters $S=3,5,7$, and $9$, when the BRI code reaches a CDF value of 1.0, is summarized in Table~\ref{Tab3}.

From Fig.\ref{figg:AverageTime}, Tables~\ref{Tab2} and~\ref{Tab3}, we observe that the proposed BRI code outperforms the baseline schemes in terms of CDC of waiting time. In particular, the BRI code is nearly unaffected by an increase in the number of stragglers. It indicates that the BRI code exhibits strong robustness against stragglers. The experimental results further demonstrate that the BRI code is able to satisfy various requirements for delay-sensitive and computation-intensive tasks in a collaborative MEC system.

As shown in Table~\ref{Tabb8}, we obtain the approximation error of our BRI code under the parameters $N=20$ and $n=8$ while varying $d$ from $0$ to $8$. It is worth noting that for $d=0$, the approximation error under this setting is identical to that of the BACC scheme~\cite{jahani2023berrut}. This is because of that the proposed BRI code becomes the BACC scheme when $d=0$. From Table~\ref{Tabb8}, we observe that the computed error achieved by the BRI code decreases as $d$ increases under the fixed parameter $n=8$. In practical scenarios, we can achieve the required approximation accuracy by adjusting parameters $d$ and $n$.

To evaluate the performance of the BRI-based linear regression (LR) algorithm, we perform experiments using the large-scale YearPredictionMSD\footnote{The download link: https://archive.ics.uci.edu/ml/datasets/YearPredictionMSD} dataset. The training and testing datasets have sizes of $(412276, 90)$ and $(103069, 90)$, respectively. As illustrated in Fig.~\ref{fig:averagetime}, we compare the average training time of the uncoded LR, EP-based LR, MatDot-based LR, LCC-based LR, block leverage score sketching-based LR~\cite{charalambides2024gradient} and BRI-based LR algorithms for training the linear regression model in a distributed system with $N=20$ and $S=3,5,7,$ and $9$. The BRI-based LR algorithm achieves the lowest average training time across scenarios $1, 2, 3,$ and $4$, while the uncoded LR algorithm exhibits the highest average training time. For the other algorithms, the average training time increases as the number of straggler nodes grows. The block leverage score sketching-based LR algorithm~ accelerates training by avoiding costly decoding steps and prioritizing important data blocks through block leverage score sampling. This outcome is mainly due to their recovery thresholds. A larger recovery threshold results in increasing the average training time as the number of straggler nodes grows. Due to its flexible recovery threshold, the BRI-based LR algorithm maintains a stable average training time regardless of the number of stragglers. The experimental results further validate the proposed BRI-based LR algorithm achieves superior performance regarding average training time compared to the other three algorithms.


\section{Conclusions}
In this paper, we have designed a novel approximate coded distributed computing scheme based on barycentric rational interpolation, called the BRI code, specific for arbitrary polynomial functions over high-dimensional datasets in a collaborative MEC system. The BRI code could obtain the final result using any returned results from workers while providing computations over both finite fields and real fields. In fact, the BRI code is able to overcome the limitations imposed on the recovery threshold. It greatly enhanced robustness against stragglers and reduced the task waiting time. Moreover, we have proposed a BRI-based gradient coding algorithm for linear regression to speed up the training process while maintaining robustness under stragglers. Finally, extensive experiments were conducted to show the superiority of the BRI code regarding average execution time and approximation error.

\bibliographystyle{IEEEtran}
\bibliography{cite}

\end{document}